\documentclass[12pt,a4paper]{article}
\usepackage[a4paper,left=2cm,right=2cm,top=3cm,bottom=2cm]{geometry}
\usepackage[utf8]{inputenc}
\usepackage[english]{babel}
\usepackage{amsmath, bbm}
\usepackage{amsfonts}
\usepackage{amssymb}
\usepackage{graphicx}
\usepackage[round]{natbib}
\usepackage{hyperref}
\usepackage{enumitem}
\usepackage{comment}
\usepackage{makecell}
\usepackage{hyperref, newfloat, caption, subcaption, natbib}
\hypersetup{
	colorlinks = true,
	urlcolor=blue,
	linkcolor=blue,
	citecolor=blue
}
\usepackage{natbib}

\newcommand{\para}{\boldsymbol{\theta} }
\newcommand{\data}{\boldsymbol{y} }
\newcommand{\dataRV}{\boldsymbol{Y} }
\newcommand{\datanoise}{\boldsymbol{\varepsilon_{\mathcal{O}}}}

\usepackage{algpseudocode}
\usepackage{algorithm}
\usepackage{multicol}

\usepackage{lineno}

\usepackage{tikz}
\usetikzlibrary{shapes,arrows}
\usetikzlibrary{positioning}
\usepackage{graphicx}
\usepackage{amsmath}
\usepackage{mathptmx} 
\usepackage{setspace}
\graphicspath{{Figures/}}

\title{An energy-based model approach to rare event probability estimation}

\author{Lea Friedli\thanks{Institute of Earth Sciences, University of Lausanne, Switzerland}
\and David Ginsbourger\thanks{Institute of Mathematical Statistics and Actuarial Science, University of Bern, Switzerland} \and Arnaud Doucet\thanks{Department of Statistics, Oxford University, United Kingdom}
\and Niklas Linde$^{\dagger}$}
\date{}

\begin{document}

\maketitle

\begin{abstract}
The estimation of rare event probabilities plays a pivotal role in diverse fields. Our aim is to determine the probability of a hazard or system failure occurring when a quantity of interest exceeds a critical value. In our approach, the distribution of the quantity of interest is represented by an energy density, characterized by a free energy function. To efficiently estimate the free energy, a bias potential is introduced. Using concepts from energy-based models (EBM), this bias potential is optimized such that the corresponding probability density function approximates a pre-defined distribution targeting the failure region of interest. Given the optimal bias potential, the free energy function and the rare event probability of interest can be determined. The approach is applicable not just in traditional rare event settings where the variable upon which the quantity of interest relies has a known distribution, but also in inversion settings where the variable follows a posterior distribution. By combining the EBM approach with a Stein discrepancy-based stopping criterion, we aim for a balanced accuracy-efficiency trade-off. Furthermore, we explore both parametric and non-parametric approaches for the bias potential, with the latter eliminating the need for choosing a particular parameterization, but depending strongly on the accuracy of the kernel density estimate used in the optimization process. Through three illustrative test cases encompassing both traditional and inversion settings, we show that the proposed EBM approach, when properly configured, (i) allows stable and efficient estimation of rare event probabilities and (ii) compares favorably against subset sampling approaches. 
\end{abstract}

Keywords: Reliability, free energy, rare event simulation, Markov chain Monte Carlo, energy-based models

\section{Introduction}
Estimating rare event probabilities is a fundamental challenge in various fields, including finance, engineering and environmental sciences. Rare events are characterized by their low occurrence rates,  but they become crucial when their outcomes have substantial consequences \citep{juneja2006rare}. In the field of uncertainty quantification, rare events are frequently related to the failure of engineering systems that are designed to be highly reliable \citep{beck2015rare}, examples of which include hydroelectric dams, airplanes, or nuclear reactors. In these cases, the accurate estimation of rare event probabilities is of utmost importance as it enables informed decision-making, effective risk management, and the design of robust systems. \\ 

In practical applications, there are commonly no analytical formulas for estimating rare event probabilities. To address this issue, asymptotic approximation methods such as the first-order reliability method (FORM; \citeauthor{hasofer1974exact} \citeyear{hasofer1974exact}) have been proposed. However, relying solely on FORM results without understanding the characteristics of the linearized domains, particularly in higher dimensions, is not recommended \citep{straub2016bayesian}. On the other hand, conventional Monte Carlo simulation methods are often computationally inefficient. Consequently, a considerable amount of research has been dedicated to developing more efficient stochastic simulation techniques for rare event probability estimations (e.g., \citeauthor{bucklew2004introduction} \citeyear{bucklew2004introduction}, \citeauthor{rubino2009rare} \citeyear{rubino2009rare}). \citet{beck2015rare} review two principal stochastic simulation approaches: importance sampling and subset simulation. Importance sampling is a variance reduction technique aiming to increase sampling frequency in the region of interest, which in the present context corresponds to the region containing the rare event and its vicinity (e.g., \citeauthor{au1999new} \citeyear{au1999new}). Subset simulation \citep{au2001estimation} represents a rare event probability as a product of larger probabilities, effectively breaking down the rare event into less rare conditional events (e.g., \citeauthor{au2014engineering} \citeyear{au2014engineering}). While subset sampling explores a wide range of parameter combinations, the more directed line sampling (e.g., \citeauthor{hohenbichler1988improvement} \citeyear{hohenbichler1988improvement}, \citeauthor{koutsourelakis2004reliability} \citeyear{koutsourelakis2004reliability}) generates samples on a hyperplane that is orthogonal to a significant direction pointing to the rare event region. \\

In this study, we are particularly interested in rare event probability estimation in the context of an underlying inverse problem. We employ non-linear Bayesian inversion aiming to infer unknown properties $\para$ given measurements $\data$. Instead of being interested in the posterior distribution itself, we target the distribution of a real-valued quantity of interest that depends on the unknown properties through a non-linear relationship $\para \mapsto \mathcal{R}(\para)$. More particularly, we seek to estimate the probability of this quantity exceeding a critical threshold, $\mathbb{P}(\mathcal{R}(\para) \geq T | \data)$, which is related to the problem of evaluating the probability of failure of a system. As the underlying relationships are non-linear, analytical formulas for the distribution of $\mathcal{R}(\para)$ conditioned on the data $\data$ are typically not available. In structural reliability engineering, similar problems have been targeted with data first being used to update $\para$ and then applying these updated distributions to the prediction of rare events (e.g., \citeauthor{papadimitriou2001updating} \citeyear{papadimitriou2001updating}, \citeauthor{jensen2013use} \citeyear{jensen2013use}, \citeauthor{sundar2013updating} \citeyear{sundar2013updating}, \citeauthor{hadjidoukas2015pi4u} \citeyear{hadjidoukas2015pi4u}). Specialized estimation methods are essential since conventional Monte Carlo approaches relying on posterior samples would typically demand overwhelmingly large sample sizes.  \\

In practice, most structural reliability methods encounter difficulties when starting with a sample approximation of the posterior, but subset sampling is an exception \citep{straub2016bayesian}. In an inversion context, it was employed to estimate the so-called ``updated robust failure probability'' by \citet{jensen2013use} and \citet{hadjidoukas2015pi4u}. Also in the engineering literature, \citet{straub2011reliability} introduced Bayesian Updating with Structural reliability methods for posterior inference (BUS; e.g. \citeauthor{straub2015bayesian} \citeyear{straub2015bayesian}). This method can be considered as an extension of rejection sampling that aims to overcome its inefficiency by treating the acceptance event as a rare event using structural reliability methods (e.g., FORM, subset simulation and line sampling; see \citeauthor{straub2016bayesian} \citeyear{straub2016bayesian}). It offers a framework, in which the rare event probability under the posterior can be estimated directly within the framework of structural reliability analysis. A recent advancement to the BUS framework involves the utilization of cross entropy-based importance sampling, which enables efficient sampling from the critical region of the posterior failure domain \citep{kanjilal2023bayesian}. \\

Free energy is a fundamental concept in materials science and physical chemistry \citep{stecher2014free}. Performing the free energy estimation on crude Monte Carlo draws is rarely practical as it would need an excessively large number of samples to sufficiently cover the whole range of the states, particularly in a rare event setting as ours. To address this challenge, one solution is to enhance the sampling in specific regions of interest by introducing a bias term. This approach, known as umbrella or non-Boltzmann sampling, can be traced back to the work of \citet{torrie1977nonphysical}, who demonstrated how to recover the unbiased probability distribution from the biased samples. Applications for free energy functions are numerous (e.g., \citeauthor{stecher2014free} \citeyear{stecher2014free}, \citeauthor{kastner2005bridging} \citeyear{kastner2005bridging}). More recently, \citet{shirts2020statistically} introduced a Bayesian formalism to estimate the free energy function by minimizing the Kullback--Leibler divergence between a continuous trial function and the empirical samples generated by biased sampling. Constructing a bias potential is a non-trivial task, therefore, \citet{valsson2014variational} introduced a variational approach that adaptively combines exploration and reconstruction by iteratively improving and refining the bias potential and the free energy function. The approach by \citet{valsson2014variational} not only utilizes the bias potential to improve the sampling technique employed for optimizing the free energy function (as with umbrella sampling), but also depends on it to directly estimate the free energy function. This variationally-enhanced sampling method has been applied in the context of coarse-graining methods by \citet{invernizzi2017coarse} and combined with machine-learning techniques by \citet{bonati2019neural}. \\

By employing concepts from energy-based models (EBMs), we propose a new formulation and approach to solve rare event probability estimation problems. We write the marginal posterior probability density function~(PDF) $r \mapsto p_{R|\dataRV}(r|\data)$ of the quantity of interest $R=\mathcal{R}(\para)$ as an energy density function with free energy $r \mapsto F(r)$, i.e. $p_{R|\dataRV}(r|\data)=\exp(-F(r))$ . To estimate $F(r)$ efficiently in the region of interest, a bias potential $r \mapsto V(r)$ and a corresponding PDF $r \mapsto p_V(r) \propto \exp(-(F(r) + V(r)))$, are introduced. That is, the energy density function of $R$ (knowing $\data$) is $r \mapsto p_{0}(r)$, where 0 refers to a zero-valued bias potential. The considered EBM (\citeauthor{goodfellow2016deep} \citeyear{goodfellow2016deep}) approach relies on optimizing the bias potential $V(r)$ such that $p_V(r)$ approximates a pre-defined PDF $r \mapsto p_{\textup{ref}}(r)$, which is selected such that it has high probability mass in the targeted region. Given the optimal bias potential $V(r)$, it is straightforward to derive the free energy $F(r)$ with an accurate estimation in the region where $p_{\textup{ref}}(r)$ has most of its mass being emphasized. Our approach is related to the variational method by \citet{valsson2014variational}, but we adopt the fundamental concepts of their method within a different formulation and apply it to a novel context, that is, rare event probability estimation. Practically, in EBM methods, the potential $V(r)$ is parameterized using methods such as neural networks, splines or radial basis functions. In \citet{valsson2014variational}, the optimal $V(r)$ is approximated by minimizing a loss function which is related to the Kullback--Leibler divergence between $p_{\textup{ref}}(r)$ and $p_V(r)$ using stochastic optimization methods. We adapt this approach and introduce a stopping criterion based on the Stein discrepancy \citep{gorham2015measuring} to achieve a balance between computational efficiency and satisfactory model performance. Furthermore, we introduce a non-parametric approach for the bias potential, thereby circumventing the difficulty of selecting an appropriate parameterization.\\

The proposed EBM approach reduces the potentially high-dimensional problem of estimating the posterior distribution of $\para$ and subsequently $\mathcal{R}(\para)$ to an optimization of a one-dimensional function $V: \mathbb{R} \rightarrow \mathbb{R}$. Importantly, the EBM approach is not limited to inversion settings and can also be effectively applied to traditional rare event probability estimation. We examine the performance of our proposed method using three illustrative test examples. The first example targets the probability of high contamination values endangering organism living in the soil. The presence of an analytical solution allows us to explore the EBM method's sensitivity to implementation variables. Next, we consider the two-dimensional four-branch function problem, commonly used as a benchmark in reliability analysis. This example does not involve inversion and demonstrates the method's suitability for general rare event estimation scenarios. Last, we use a simple load and capacity example from \citet{straub2016bayesian} to compare the EBM method's performance against their BUS approach. The manuscript is organized as follows: Section~\ref{EBM_ebm_metho} introduces the problem setting, outlines our energy-based model approach, and details the methods employed for comparison. In Section~\ref{EBM_ebm_studies}, the test examples are presented. Subsequently, Section~\ref{EBM_ebm_disc} discusses the results, leading to the conclusions exposed in Section~\ref{EBM_ebm_conc}.\\

\section{Methodology}
\label{EBM_ebm_metho}

\subsection{Problem setting}
\label{EBM_ebm_setting}
In the considered rare event setting, we target a quantity of interest $R = \mathcal{R}(\para)$ derived from the random vector $\para$ on $\mathbb{R}^d$ by some non-linear function $\mathcal{R}: \mathbb{R}^d \rightarrow \mathbb{R}$. We consider a rare set $A = \{ \para \in \mathbb{R}^d: \mathcal{R}(\para) \geq T \}$ with $T \in \mathbb{R}$ and want to estimate $\mathbb{P}(\para \in A)$. In a traditional rare event setting, we consider $\para$ distributed according to a prior PDF~$p_{\para}(\para)$, which is absolutely continuous with respect to a dominating measure, typically the Lebesgue measure on $\mathbb{R}^d$. The rare event probability can then be expressed as, 
\begin{equation}
    \mathbb{P}(\mathcal{R}(\para) \geq T) = \mathbb{P}(\para \in A) = \int_A p_{\para}(\para) \mathrm{d} \para.
\end{equation}

Bayesian probabilistic inversion methods target generally unknown properties $\para$ and seek to infer their posterior PDF given the measurements $\data \in \mathbb{R}^m$. In most applications, the random data vector is given by $\dataRV = \mathcal{G}(\para) + \datanoise$, with $\mathcal{G}: \mathbb{R}^d \rightarrow \mathbb{R}^m$ referring to the forward operator and $\datanoise$ to the observational noise. In Bayes' theorem, the posterior PDF of the target parameters $\para$ given measurements $\data$ is given by, 
\begin{align}
	p_{\para|\dataRV}(\para | \data ) = \frac{p_{\para}(\para) p_{\dataRV|\para}(\data | \para) }{p_{\dataRV}(\data)},
\end{align}
with the prior PDF~$p_{\para}(\para)$, the likelihood function~$p_{\dataRV|\para}(\data | \para)$ and the evidence~$p_{\dataRV}(\data)$ (assumed non-zero). Assuming independent Gaussian observational errors, the likelihood is expressed as $p_{\dataRV|\para}(\data | \para)=\varphi_{m}(\data; \mathcal{G}(\para), \boldsymbol{\Sigma_{\dataRV}})$, with $\varphi_{m}(\cdot; \boldsymbol{\mu}, \boldsymbol{\Sigma})$ denoting the PDF of a $m$-variate normal distribution with mean $\boldsymbol{\mu}$ and diagonal covariance matrix $\boldsymbol{\Sigma}$ with the variance of the observational error on its diagonal. For $\para$ distributed according to a Bayesian posterior PDF $\para \mapsto p_{\para|\dataRV}(\para|\data)$, we aim to estimate the rare event probability,
\begin{equation}
\label{EBM_prob_eq}
    \mathbb{P}(\mathcal{R}(\para) \geq T| \data) = \mathbb{P}(\para \in A | \data) = \int_A p_{\para|\dataRV}(\para|\data) \mathrm{d} \para.
\end{equation}

\subsection{Energy-based model approach}
\label{EBM_EBM}
While our methodology is formulated for an inversion setting, it can be readily adjusted to suit a conventional rare event estimation scenario with $\para$ being distributed according to a prior PDF. \\

We write the posterior PDF of $\para$ as an energy density function,
\begin{equation}
    p_{\para|\dataRV}(\para | \data) = \frac{\exp(-U(\para))}{Z}, \quad Z := \int \exp(-U(\zeta))\mathrm{d} \zeta, 
\end{equation}
with free energy $U(\para) = - \log p_{\dataRV|\para}(\data |\para) - \log p_{\para}(\para)$. The marginal posterior distribution of the quantity of interest $R=\mathcal{R}(\para)$ is given by,
\begin{equation}
    p_{R|\dataRV}(r|\data) = \int  \frac{\exp(-U(\para))}{Z} \delta (r - \mathcal{R}(\para)) \mathrm{d} \para.
\end{equation}
More details on the definition of such integrals can be found in \citet{stoltz2010free} (Lemma~3.2; co-area formula). We write $p_{R|\dataRV}(r|\data)$ as an energy density function with (unknown) free energy $r \mapsto F(r)$,
\begin{equation}
\label{postR}
    p_{R|\dataRV}(r|\data) = \exp(-F(r)). 
\end{equation}
In EBMs, we aim to approximate this free energy $F(r)$. This function can be parameterized with any nonlinear regression function, effectively transforming density estimation into a nonlinear regression problem \citep{song2021train}. While the exponential family of distributions, for example, imposes stronger assumptions on the form of the PDF, an EBM only necessitates the PDF to be non-zero within its domain.\\

\begin{figure}[h]
\centering
    \includegraphics[width=.9\textwidth]{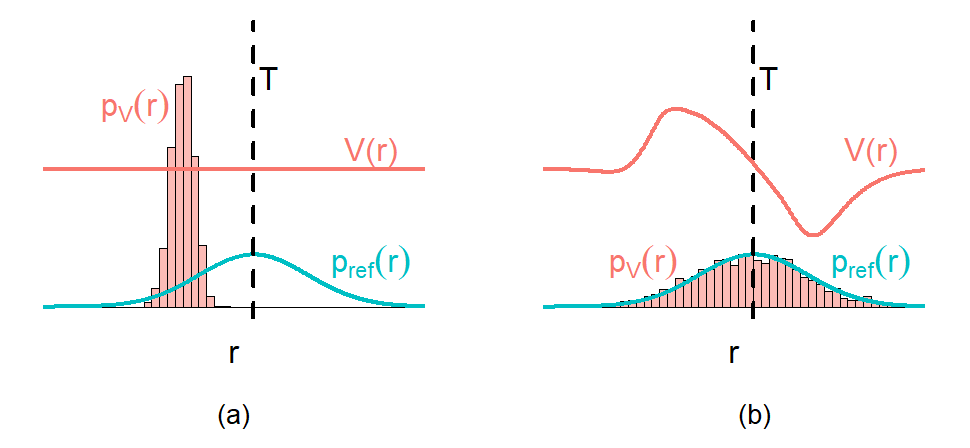}
	\caption{
		Illustration of the energy-based model approach for rare event probability estimation. (a) For $V(r)=0$, $p_V(r) = p_{R|\dataRV}(r|\data)$ has low probability mass on the set $\{R \geq T \}$. (b) For $V(r) \approx V_{\textup{opt}}(r)$, $p_V(r) \approx p_{\textup{ref}}(r)$ yielding by design a high probability mass on the set $\{R \geq T \}$.}
	\label{EBM_fig:illu_ebm}
\end{figure}

In order to estimate $F(r)$ and $p_{R|\dataRV}(r|\data)$, we have the option of utilizing posterior samples obtained from $p_{\para|\dataRV}(\para | \data)$ and applying a transformation $\para \mapsto \mathcal{R}(\para)$. Since this follows a typical Monte Carlo approach, it would need an impractically large number of samples to include the low probability regions of $p_{R|\dataRV}(r|\data)$ that are of interest. To enhance sampling in the region of interest, we seek to sample in accordance to a pre-defined $r \mapsto p_{\textup{ref}}(r)$ with substantial mass on the set $\{R \geq T \}$. To achieve this, we introduce the bias potential $r \mapsto V(r)$ and the PDF $r \mapsto p_V(r)$,
\begin{equation}
\label{pvR}
    p_V(r) = \frac{\exp(-(F(r) + V(r)))}{\int \exp(-(F(s) + V(s))) \mathrm{d} s}.
\end{equation} 
We seek $p_V(r)=p_{\textup{ref}}(r)$, which holds for $r \mapsto V_{\textup{opt}}(r) = -F(r) - \log(p_{\textup{ref}}(r))$ (for $p_{\textup{ref}}(r) > 0$, ignoring an irrelevant constant). To estimate this optimal bias potential is the primary objective of the presented EBM approach. Given $V_{\textup{opt}}(r)$, it is then straightforward to obtain $F(r)$ as $F(r) = - \log(p_{\textup{ref}}(r)) - V_{\textup{opt}}(r)$, for $p_{\textup{ref}}(r) > 0$ and hence $p_{R|\dataRV}(r|\data)$. Eventually, we can express the probability of the rare event as,
\begin{equation}
\label{rareprob}
    \mathbb{P}(\mathcal{R}(\para) \geq T | \data) = \int_{T}^{\infty} p_{R|\dataRV}(r|\data) \mathrm{d} r = \int_{T}^{\infty} \exp(-F(r)) \mathrm{d} r.
\end{equation}
A basic visualization of the EBM approach is given in Figure~\ref{EBM_fig:illu_ebm}. Figure~\ref{EBM_fig:illu_ebm}(a) shows the initialization of the method with a zero bias potential $V(r)=0$ and correspondingly $p_V(r) = p_{R|\dataRV}(r|\data)$ with low probability mass on the set $\{R \geq T \}$. Figure~\ref{EBM_fig:illu_ebm}(b) shows the optimized bias potential $V(r) \approx V_{\textup{opt}}(r)$ and the corresponding $p_V(r) \approx p_{\textup{ref}}(r)$, which now has high probability mass on the set $\{R \geq T \}$. Intuitively, one could imagine that the optimized bias potential describes how the free energy surface must be distorted in order to shift $p_{R|\dataRV}(r|\data)$ to $p_{\textup{ref}}(r)$. \\

An important choice is the definition of $p_{\textup{ref}}(r)$. When the PDF $p_{R|\dataRV}(r|\data)$ is supported by a compact interval with length $\Omega_p$, one possible and natural choice is to set $p_{\textup{ref}}(r) = 1/\Omega_p$ \citep{valsson2014variational}. This results in uniform sampling and is commonly employed in other enhanced sampling approaches \citep{wang2001efficient}. If the support is unbounded, then $p_{\textup{ref}}(r)$ can be selected such that it shifts mass to the range of $R$ which is of interest (here $R \geq T$). To enable accurate rare event probability estimation, the right tails of the selected $p_{\textup{ref}}(r)$ have to be at least as heavy as the tails of the target PDF $p_{R|\dataRV}(r|\data)$. \\

\subsubsection{Parameterization and optimization of the bias potential}
We parameterize the one-dimensional bias potential $r \mapsto V(r)$ using a flexible model $V_{\psi}(r)$ with free parameters $\psi$. We employ here radial basis functions (RBFs) with squared exponential kernels,
\begin{equation}
    \label{EBM_RBF}
    V_{\psi}(r) = \sum_{j=1}^{B} w_j \phi(r-b_j), \text{ where } \phi(r) = \exp(-(\kappa_j r)^2).
\end{equation}
For a number $B$ of RBFs, we obtain $\psi = (\mathbf{w, b, \kappa}) \in \mathbb{R}^B \times \mathbb{R}^B \times \mathbb{R}^B $. Then, we seek the optimal $V_{\psi}(r)$ by minimizing the Kullback--Leibler divergence between $p_{\textup{ref}}(r)$ and $p_{V_\psi}(r)$ with respect to $\psi$. We achieve this by employing stochastic gradient descent with momentum (SGDM; e.g., \citeauthor{liu2020improved} \citeyear{liu2020improved}),
\begin{align}
\label{sgdm}
    m_{n} &= \beta m_{n-1} + (1-\beta)  \frac{\partial \textup{KL}(p_{\textup{ref}}||p_{V_\psi})}{\partial \psi} , \\
    \psi_{n+1} &= \psi_n - \gamma m_n, \nonumber
\end{align}
using initial momentum $m_0=0$, momentum weight $\beta$, learning rate $\gamma$ and a stochastic gradient of $\textup{KL}(p_{\textup{ref}}||p_{V_\psi})$ at $\psi_n$. For a momentum weight of $\beta=0$, SDGM reduces to traditional stochastic gradient descent. \\

It holds that
\begin{align}
\label{KLdiv}
    \textup{KL}(p_{\textup{ref}}||p_{V_\psi}) &= \int p_{\textup{ref}}(x) \log p_{\textup{ref}}(x) \mathrm{d}x - \int p_{\textup{ref}}(x) \log p_{V_\psi}(x) \mathrm{d}x \\ 
    &= \textup{const.} + \int p_{\textup{ref}}(x) V_{\psi}(x) \mathrm{d}x +    \log \int \exp(-(F(s) + V_{\psi}(s))) \mathrm{d} s, \nonumber
\end{align} 
and (assuming standard regularity assumptions that allow swapping the order of differentiation and integration),
\begin{align}
     \frac{\partial \textup{KL}(p_{\textup{ref}}||p_{V_\psi})}{\partial \psi} &=
     \frac{\partial}{\partial \psi} \int p_{\textup{ref}}(x) V_{\psi}(x) \mathrm{d}x + \frac{\partial}{\partial \psi}   \log \int \exp(-(F(s) + V_{\psi}(s))) \mathrm{d}s \\
      &=  \int p_{\textup{ref}}(x) \frac{\partial}{\partial \psi} V_{\psi}(x) \mathrm{d}x -
      \int p_{V_{\psi}}(s) \frac{\partial}{\partial \psi} V_{\psi}(s) \mathrm{d}s. \nonumber
\end{align}
This allows us to approximate,
\begin{equation}
\label{EBM_loss_n}
    \frac{\partial \textup{KL}(p_{\textup{ref}}||p_{V_\psi})}{\partial \psi} \approx \frac{1}{n} \sum_{i=1}^{n}\frac{\partial}{\partial \psi }V_{\psi}(X_i) - \frac{1}{n} \sum_{i=1}^{n}\frac{\partial}{\partial \psi }V_{\psi}(R_i), \quad \text{ with } X_i \sim p_{\textup{ref}}(\cdot), R_i \sim p_{V_{\psi}}(\cdot).
\end{equation}
This requires samples from $p_{V_{\psi}}(\cdot)$ and $p_{\textup{ref}}(\cdot)$, with the latter being straightforward as the PDF $p_{\textup{ref}}(\cdot)$ is pre-defined. Sampling from $p_{V_{\psi}}(\cdot)$ directly is not possible as it would require that the free energy $F(\cdot)$ was known. Thus, we use samples of the original space of $\para$ by introducing, 
\begin{equation}
\label{EBM_pv_theta}
    p_{V_\psi}(\para) = \frac{\exp(-(U(\para)+ V_{\psi}(\mathcal{R}(\para))))}{\int \exp(-(U(\boldsymbol{\zeta})+ V_{\psi}(\mathcal{R}(\boldsymbol{\zeta})))) \mathrm{d} \boldsymbol{\zeta}},
\end{equation}
and then transform them using $\mathcal{R}(\cdot)$. This is correct as one can easily check that the following identity $\int p_{V_\psi}(\para) \delta(r-\mathcal{R}(\para)) \mathrm{d} \para = p_{V_\psi}(r)$ holds. In our implementation, the samples of $p_{V_\psi}(\para)$ are obtained using a Metropolis--Hastings algorithm (MH; \citeauthor{metropolis1953equation} \citeyear{metropolis1953equation}; \citeauthor{hastings1970monte} \citeyear{hastings1970monte}). \\

This idea of optimizing the bias potential to estimate the free energy coincides with the variational approach of \citet{valsson2014variational}, \citet{invernizzi2017coarse} and \citet{bonati2019neural} introduced in metadynamics, working with the functional,
\begin{equation}
\label{eq_func}
    \Omega(V) = \log \left( \frac{ \int \exp(-(F(x) + V(x))) \mathrm{d} x}{\int \exp(-F(x)) \mathrm{d} x} \right) + \int p_{\textup{ref}}(x) V(x) \mathrm{d} x,
\end{equation}
which they show is convex and invariant under the addition of an arbitrary constant to the bias potential. It is easy to show that minimizing $\Omega(V)$ is equivalent to minimizing $\textup{KL}(p_{\textup{ref}}||p_V)$ (e.g. \citeauthor{invernizzi2017coarse} \citeyear{invernizzi2017coarse}). Furthermore, we can reformulate the minimization of the Kullback--Leibler divergence as a maximum likelihood estimation problem (Appendix~\ref{EBM_app:MLE}), which offers multiple theoretical possibilities as it allows us to transfer known theoretical results on maximum likelihood estimation to our approach. \\

\subsubsection{Non-parametric bias potential}
\label{nonparV}
In addition to the parametric bias potentials $V_{\psi}$ considered so far, the functional $\Omega(V)$ of Eq.~(\ref{eq_func}) can be applied to a broader class of bias potentials $V$. While \citeauthor{invernizzi2017coarse} \citeyear{invernizzi2017coarse} is not providing details regarding the source space on which $\Omega(V)$ is defined, it is noticeable that for $V$ integrable with respect to $p_{\text{ref}}$, $\Omega(V)$ can be seen as a convex functional (taking infinite values in case of non-integrable $\exp(-F-V))$. Furthermore, for suitable $V$ and directions $H$, one can work out directional derivatives of $\Omega$ 
\begin{equation}
    \mathrm{d}\Omega(V;H) = \int p_{\textup{ref}}(x) H(x) \mathrm{d}x -
      \int p_{V}(x) H(x) \mathrm{d}x=\int (p_{\textup{ref}}(x) - p_{V}(x)) H(x) \mathrm{d}x. 
\end{equation}
In particular, for bounded $V$ and $p_{\text{ref}}$ as considered herein, the directional derivative at $V$ in the direction $p_{\textup{ref}} - p_{V}$ is well defined and $p_{V}-p_{\textup{ref}}$ delivers a descent direction. In sequential settings, the function $V_n$ can be updated with SGDM (Eq. \ref{sgdm}) using $p_{\textup{ref}} - p_{V}$ in lieu of stochastic gradient at $V_n$. While for the PDF $p_{\textup{ref}}$, we have access to its analytical form, we rely here in practice for $p_{V_{n}}$ on a kernel density estimate derived from the MH samples. \\

\subsubsection{Stopping criteria}
\label{EBM_stop}
Appropriate stopping criteria are crucial when employing stochastic gradient-based optimization in order to strike a balance between computational efficiency and achieving satisfactory model performance. One naïve approach is to stop the training after a pre-defined number of optimization steps. Since this approach does not ensure convergence if the selected number is too low or might waste computational resources if chosen to be too high, it makes sense to monitor the values of the loss function and terminate the optimization when it reaches a desired criterion or when it no longer improves significantly. \\

Here we consider a convergence measure based on Stein's method \citep{stein1972bound}: the Stein discrepancy \citep{gorham2015measuring}. To circumvent the latter's computational intractability, we use a specific version known as the kernel Stein discrepancy (KSD; \citeauthor{liu2016kernelized} \citeyear{liu2016kernelized}, \citeauthor{chwialkowski2016kernel} \citeyear{chwialkowski2016kernel}, \citeauthor{gorham2017measuring} \citeyear{gorham2017measuring}),
\begin{align}
\label{KSD}
    \textup{KSD}(p_{\textup{ref}}||p_{V_\psi}) =& \sqrt{\frac{1}{n^2} \sum_{i,j=1}^{n} k_{p_{\textup{ref}}} (R_i, R_j)}, \quad R_i \sim p_{V_\psi}(\cdot), \text{  with,  } \\
    k_{p_{\textup{ref}}} (r,s) =& \nabla_{r} \nabla_{s} k(r,s) + \langle \nabla_{r} k(r,s), \nabla_s \log p_{\textup{ref}}(s) \rangle + \langle \nabla_{s} k(r,s), \nabla_r \log p_{\textup{ref}}(r) \rangle \nonumber \\
    &+ k(r,s) \langle \nabla_{r} \log p_{\textup{ref}}(r), \nabla_s \log p_{\textup{ref}}(s) \rangle, \nonumber
\end{align}
with kernel $k(r,s)$. The use of the KSD is particularly advantageous in our setting due to the analytical knowledge about the probability distribution $p_{\textup{ref}}(\cdot)$. \\

As stopping criterion, we employ a statistical test for goodness-of-fit based on the squared KSD (\citeauthor{liu2016kernelized} \citeyear{liu2016kernelized}, \citeauthor{chwialkowski2016kernel} \citeyear{chwialkowski2016kernel}). Employing bootstrap sampling for the estimation of the test statistic, the algorithm stops as soon as the null hypothesis of $p_{\textup{ref}} = p_{V_\psi}$ cannot be rejected with a significance level of $\alpha$. To counteract the conservative nature of the test procedure when dealing with the correlated samples resulting from MH, \citet{chwialkowski2016kernel} suggest an approach based on the wild bootstrap technique (e.g., \citeauthor{shao2010dependent} \citeyear{shao2010dependent}). The proposed test statistics,
\begin{equation}
\label{EBM_GOF_corr}
    \frac{1}{n^2} \sum_{i,j=1}^n W_i W_j k_{p_{\textup{ref}}}(R_i,R_j), \quad \text{using} \quad
    W_i = \mathbbm{1}(U_i > a_{BS}) W_{i-1} - \mathbbm{1}(U_i < a_{BS}) W_{i-1},
\end{equation}
takes into account the correlation structure in the samples by mimicking it with an auxiliary Markov chain taking values in $\{-1,1\}$, where $W_1=1$ and $U_i \sim \textup{Unif}(0,1)$. \citet{chwialkowski2016kernel} suggest combining this method with thinning of the generated samples. They recommend to thin the chain such that $\textup{Cor}(R_i,R_{i-1})~<~0.5$, set $a_{BS} = 0.1/q$ with $q<10$ and use at least $n = \max(500q,100)$ samples. A workflow of the full EBM method employing a parametric form for the bias potential is depicted in Figure \ref{fig:flow}. \\

\begin{figure}
	\centering
    \tikzstyle{block} = [rectangle, draw, fill=black!10, text width=20em, text centered, rounded corners, node distance=3cm]
    \tikzstyle{line} = [draw, -latex']
	\begin{tikzpicture}[minimum size=5mm,
    node distance=4cm and 7cm,
    >=stealth,
    bend angle=45,
    auto]
    \node [block] (zero) {\begin{minipage}{\linewidth}
      \vspace{0.25cm} Choose parametric form of bias potential $V_{\psi}(r)$ \\ and initialize parameters $\psi$ such that $V_{\psi}(r) \equiv 0$ \end{minipage}};
    \tikzstyle{block} = [rectangle, draw, fill=white, text width=20em, text centered, rounded corners, node distance=3cm]
    \node [block, below=1cm of zero] (one) {\begin{minipage}{\linewidth}\vspace{0.35cm} \noindent Use MH to sample from $p_{V_\psi}(\para)$ (Eq. \ref{EBM_pv_theta})\\ and transform samples using $\mathcal{R}(\para)$ \end{minipage} };
    \node [block, below=1cm of one] (one2) {\begin{minipage}{\linewidth}
      \vspace{0.25cm}
      Based on samples $R_i \sim p_{V_\psi}(\cdot)$, check if \\
    $p_{\textup{ref}}~=~p_{V_\psi}$ can be rejected (Section \ref{EBM_stop})
    \end{minipage} };    
    \tikzstyle{block} = [rectangle, draw, fill=white, text width=14em, text centered, rounded corners, node distance=3cm]
    \node [block, right=1.5cm of one2] (one1) 
    {\begin{minipage}{\linewidth}
      \vspace{0.35cm} 
    Employ SGDM (Eqs. \ref{sgdm}, \ref{EBM_loss_n}) to update the parameters $\psi$   \end{minipage} };
    \tikzstyle{block} = [rectangle, draw, fill=white, text width=20em, text centered, rounded corners, node distance=3cm]
    \node [block, below=1cm of one2, fill=black!10, text opacity=1] (two) 
    {\begin{minipage}{\linewidth}
      \vspace{0.35cm} Use optimized $V_\psi(r)$ to estimate the rare event  \\ probability $\mathbb{P}(\mathcal{R}(\para) \geq T | \data)$ (Eq. \ref{rareprob})
    \end{minipage}  };
    \tikzstyle{fancytitle} =[fill=blue!30, text=black]
    \node[fancytitle, right=10pt] at (zero.north west) {\textbf{Initialization}};
    \node[fancytitle, right=10pt] at (one.north west) {\textbf{Sample $R_i\sim p_{V_\psi}(\cdot)$}};
    \node[fancytitle, right=10pt] at (one2.north west) {\textbf{Stopping criterion}};
    \node[fancytitle, right=10pt] at (two.north west) {\textbf{Completion}};
    \node[fancytitle, right=10pt] at (one1.north west) {\textbf{Update $V_{\psi}(r)$}};
    \path [line, very thick] (zero) -- (one) ;
    \path [line, very thick] (one) -- (one2) ;
    \path [line, very thick] (one2) -- node[anchor=north] {YES} (one1);
    \path [line, very thick] (one2) -- node[anchor=east] {NO} (two);
   \path [draw, very thick, ->] (one1.north) ++(0.5,0) -- ++(0,3.2) -- ++(-7,0) -- ++(0,-0.7);
\end{tikzpicture}
	\caption{Work flow of the EBM method using a parametric form for the bias potential $V_{\psi}(r)$. }
	\label{fig:flow}
\end{figure}

\subsubsection{Distinction from classical importance sampling}
Importance sampling (IS) encompasses a set of Monte Carlo techniques where the expectation of the image  of a random variable $R$ via a prescribed function $f$, say $\mathcal{I}_f = \mathbb{E}_p[f(R)]$, is estimated using a weighted average of $f(R_i)$ $(i=1\dots,n)$ where the $R_i$'s form an sample from another instrumental distribution \citep[e.g.,][]{tokdar2010importance}. 
Here both the original distribution and the instrumental distribution are assumed to be absolutely continuous with respect to a common dominating measure, with respective densities $p$ and $\pi$. 
Estimating $\mathcal{I}_f$ may in fact be particularly challenging when a significant portion of $f$'s variability arises from regions of $R$'s domain having low probability under $p$ \citep{owen2000safe}. 
The rationale of IS is thus to introduce an easy-to-sample-from $\pi$ (also called the IS density) that samples frequently from such regions. For any PDF $r \mapsto \pi_R(r)$ satisfying $\pi_R(r)>0$ whenever $f(r)p(r)\neq0$, it holds that $\mathcal{I}_f = \mathbb{E}_{\pi}[f(R)\frac{p(R)}{\pi_R(R)}]$ \citep{tokdar2010importance}. Accordingly, replicates $R_1,...,R_N \sim \pi_R(\cdot)$ can be used to estimate $\mathcal{I}_f$ by,
\begin{equation}
    \hat{\mathcal{I}_f} = \frac{1}{N} \sum_{i=1}^N f(R_i) \frac{p(R_i)}{\pi_R(R_i)}.
\end{equation}
In our setting (Section \ref{EBM_ebm_setting}), the rare event probability for an inversion problem is given by $\mathcal{I}_f$ with $f(r)~=~\mathbbm{1}(r~\geq~T)$ and $p(r)=p_{R|\dataRV}(r|\data)$ (Eq.~\ref{rareprob}), and the form of a corresponding IS estimator would be,
\begin{equation}
\label{IS}
    \mathbb{P}(R \geq T |\data) \; \hat{=} \frac{1}{N} \sum_{i=1}^N \mathbbm{1}(R_i \geq T) \frac{p_{R|\dataRV}(R_i|\data)}{\pi_R(R_i)}, 
\end{equation}
with samples from the IS density $R_i \sim \pi_R(\cdot)$. In our setting, $p_{R|\dataRV}(r|\data)$ can generally not be evaluated as $F$ is not known. Therefore, IS has to rely on sampling of the original space of $\para$ with,
\begin{equation}
\label{IS_para}
    \mathbb{P}(R \geq T |\data) \; \hat{=} \frac{1}{N} \sum_{i=1}^N \mathbbm{1}(\mathcal{R}(\para_i) \geq T) \frac{p_{\para|\dataRV}(\para_i|\data)}{\pi_{\para}(\para_i)}, \quad \para_i \sim \pi_{\para}(\cdot).
\end{equation}
Since we know the unnormalized posterior $p_{\para|\dataRV}(\para | \data)$, a self-normalized version of IS can be applied \citep{tokdar2010importance}. An optimal IS density $\pi_{\para}(\cdot)$, which minimizes the variance of the estimator, is proportional to $\mathbbm{1}(\mathcal{R}(\para)~\geq~T)~p_{\para|\dataRV}(\para|\data)$ \citep{kahn1953methods}. In practice, this means identifying an IS density that frequently samples $\para$ such that $\mathcal{R}(\para)$ lies in the region of the rare event threshold (while still accounting for the data $\data$). \\ 

Let us now contrast this with the presented EBM approach to rare event probability estimation. Here, the rare event probability is estimated by integrating over an approximation of the posterior PDF $p_{R|\dataRV}(\cdot|\data)$,
\begin{equation}
    \mathbb{P}(R \geq T | \data) = \int_{T}^{\infty} p_{R|\dataRV}(r|\data) \mathrm{d} r \; \approx\int_{T}^{\infty} \exp(-\hat{F}(r)) \mathrm{d} r,
\end{equation}
with plugin in $\hat{F}(r) = - \log(p_{\textup{ref}}(r)) - V(r)$ using the optimized bias potential $V(r) \approx V_{\textup{opt}}(r)$ (see also Eq.~\ref{rareprob}). The only step where the EBM approach employs sampling is in the SGDM used to optimize the bias potential (Eq. \ref{EBM_loss_n}). However, while the presented EBM approach estimates the rare event probability in a entirely different way, it could be applied within a classical IS approach to identify a suitable IS density $\pi_{\para}(\cdot)$ for Equation~\ref{IS_para}. Since the EBM method optimizes the bias potential in a way that the transformed samples from $p_{V}(\para)$ (Eq. \ref{EBM_pv_theta}) coincide with $p_{\textup{ref}}(\cdot)$ (which has significant mass in the region around the rare event), this provides a promising IS density. This IS option is not pursued in the present study. \\

\subsubsection{Related work in machine learning}
Energy-based models are a popular class of probabilistic models in machine learning \citep{goodfellow2016deep}. In this context, EBMs are used to model the distribution of available data $(r_i)_{i=1}^N$ (say images) distributed according to an unknown PDF $p_{\textup{ref}}(\cdot)$ via a model of the form 
\begin{equation}
p_{\psi}(r)=\frac{\exp(-U_\psi(r))}{Z_\psi},
\end{equation}
the potential $U_\psi(r)$ being typically parameterized by a neural network. The parameter $\psi$ can be learned by maximizing the normalized log-likelihood, that is,
\begin{equation}
\ell_N(\psi)=\frac{1}{N} \sum_{i=1}^N\log p_\psi(r_i),
\end{equation}
which approximates, up to an additive constant independent of $\psi$, the Kullback--Leibler divergence $\textup{KL}(p_{\textup{ref}}||p_\psi)$ as $N \rightarrow \infty$ (Appendix \ref{EBM_app:MLE}). Similarly to our scenario, the gradient of this discrepancy requires sampling from $p_\psi(\cdot)$ using Markov chain Monte Carlo. \\

Compared to applications in machine learning, the rare event simulation context discussed in this paper enjoys several attractive properties. First, we can obtain as many samples from $p_{\textup{ref}}(r)$ as we want. Second, because we know $p_{\textup{ref}}(r)$ analytically, we can assess how good our estimate of $p_{\textup{ref}}(r)$ is via $p_\psi(r)$ as discussed in Section~\ref{EBM_stop}. Third, we only need to estimate a one-dimensional function $V_{\psi}(r)$ instead of a complex high-dimensional potential $U_\psi(r)$. We note finally that numerous alternatives to Kullback--Leibler minimization have been developed to train EBMs and that some of them might be applicable to the rare event simulation context, see \citet{song2021train} for a recent review.\\

\subsection{Alternative rare event probability estimation methods}
\label{EBM_alternatives}

\subsubsection{Subset sampling}
\label{EBM_MH_SS}
Subset sampling for rare event probability estimation was introduced by \citet{au2001estimation} and is based on expressing the probability of a rare event as a multiplication of higher conditional probabilities involving intermediate failure events. To estimate $\mathbb{P}(\para \in A | \data) = \mathbb{P}(\mathcal{R}(\para) \geq T | \data)$, subset sampling employs a sequence of increasing thresholds $\{ T_0, ...,T_K \}$ with $T_0=-\infty$ and $T_K = T$ and corresponding nested sets $A_k = \{ \para \in \mathbb{R}^d: \mathcal{R}(\para) \geq T_k \}$. It holds that, 
\begin{equation}
    \mathbb{P}(\para \in A | \data) = \prod\limits_{k=1}^{K} \mathbb{P} (\para \in A_{k} | \para \in A_{k-1}, \data).
\end{equation}
\citet{au2001estimation} propose to select the sequence of thresholds adaptively such that a ratio of $(1-\gamma)$ of the samples survive. Such an adaptive procedure leads to positively biased estimates, however, this bias decreases with an increasing number of samples per subset \citep{cerou2012sequential}. To use subset sampling in our inversion setting, we initialize the first set of samples with a draw from the posterior using the states sampled by a MH algorithm after burn-in. In this context, the thinning factor is critical as the initial set of samples should be as independent as possible. We then propagate the set of samples to estimate the conditional probabilities by sampling from the subsets while accounting for the posterior. A similar approach was applied by \citet{jensen2013use}, who rely on so-called transitional MCMC to generate the first set of posterior samples.\\

\subsubsection{Bayesian updating of rare event probabilities}
\label{EBM_BUS}
In \citep{straub2015bayesian}, building upon \citep{straub2011reliability}, a methodological framework called Bayesian Updating with Structural reliability methods (BUS) is introduced. This method serves as an extension of the conventional rejection sampling technique in Bayesian analysis. In a basic version of rejection sampling, samples are generated from the prior distribution and subsequently accepted with a probability of $p_{\dataRV|\para}(\data | \para)/\bar{p}$, where $\bar{p}$ is an upper bound on $\mathrm{sup}_{\para} p_{\dataRV|\para}(\data | \para)$. The issue with the basic rejection sampling algorithm is its inefficiency, particularly when dealing with high-dimensional or complex posterior distributions. The core concept behind BUS is to address the challenge posed by the small acceptance probabilities in rejection sampling with structural reliability methods.  \\

We consider BUS with subset sampling, employing the principles of subset sampling (Section~\ref{EBM_MH_SS}) to sample from the acceptance region of the extended rejection sampling algorithm, while enabling an adaptive estimation of the constant $\bar{p}$ \citep{betz2014adaptive}. In the study of \citet{straub2016bayesian}, this method strikes an appropriate balance between accuracy and the number of model evaluations and has the advantage of being more robust compared to line sampling that they also consider. To apply BUS for rare event probability estimation in an inversion setting, \citet{straub2016bayesian} combine two subset sampling runs, where the first generates samples from the posterior and the second propagates these samples towards the targeted rare event. This approach is comparable to the subset sampling method described in Section~\ref{EBM_MH_SS} with the difference that the posterior sampling part is replaced with BUS. \\

\section{Illustrative test examples}
\label{EBM_ebm_studies}
We now explore the performance of our EBM method in different illustrative test examples. We start with exploring the method's sensitivity to the choice of configuration using a hypothetical contamination problem for which an analytical solution is available. Next, we consider the two-dimensional four-branch function problem, a frequently utilized benchmark in reliability analysis. This example does not require inversion and effectively demonstrates the method's suitability for more traditional rare event estimation scenarios. For both of these first examples, we compare the EBM method against a standard implementation of subset sampling. Finally, we consider a simple load and capacity example of \citet{straub2016bayesian} to compare the EBM method's performance against the BUS approach with subset sampling. \\

\begin{figure}
    \begin{subfigure}[b]{.32\textwidth}
         \centering
         \includegraphics[width=5cm]{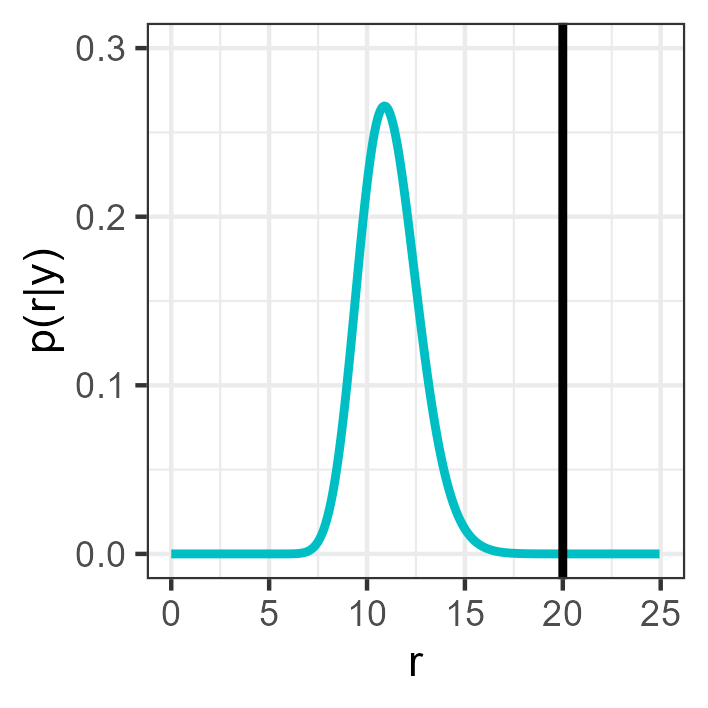}
         \caption{}
         \label{EBM_fig:contamin_postR}
     \end{subfigure}
     \begin{subfigure}[b]{.32\textwidth}
         \centering
         \includegraphics[width=5cm]{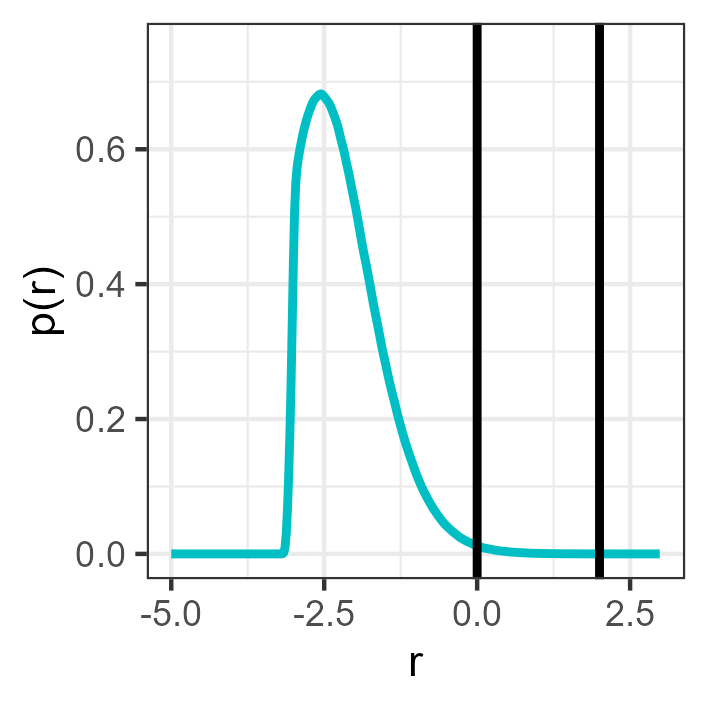}
         \caption{}
         \label{EBM_fig:branch_postR}
     \end{subfigure}
     \begin{subfigure}[b]{.32\textwidth}
         \centering
         \includegraphics[width=5cm]{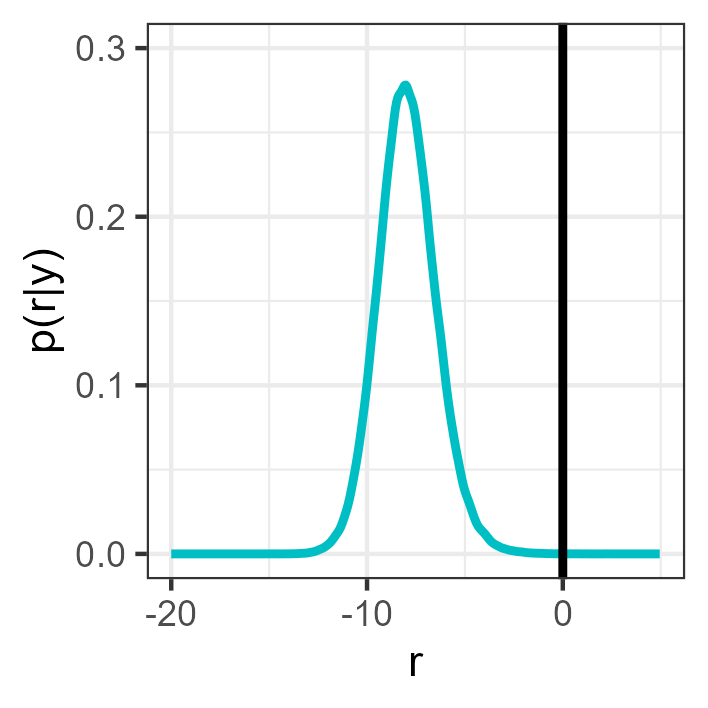}
         \caption{}
        \label{EBM_fig:load_postR}
     \end{subfigure}
		\caption{PDFs of the quantity of interest $R$ in the three illustrative examples: (a) The dose response in Section~\ref{EBM_ebm_analytical}, (b) the four-branch function in Section~\ref{EBM_ebm_branch} and (c) the difference between load and capacity in Section~\ref{EBM_ebm_loadcapa} ($n_C=10$). The vertical lines show the different critical thresholds $T$ targeted within $\mathbb{P}(\mathcal{R}(\para) \geq T | \data)$. }
		\label{EBM_fig:qoi_post}				
\end{figure}

\subsection{Analytical contamination example}
\label{EBM_ebm_analytical}
In our first test example, we seek to infer a hypothetical contamination field $\para$ using cell measurements~$\data$ and aim at predicting the probability of a resulting dose response $\mathcal{R}(\para)$ being critically high for organisms living in the soil. We consider a setting with $M=9$ contamination cells. The ``true" values of the contamination level are sampled using independent Gaussian PDFs $\theta_m \sim \mathcal{N}(1, 0.3^2)$. For the data $\data$, we measure the concentration of the contamination in three cells and assume independent Gaussian observation errors with a standard deviation of $0.05$. We target the quantity of interest $\mathcal{R}(\para) = \sum_{m=1}^M \theta_m^2$, which is related to the linear quadratic dose response (e.g., \citeauthor{mcmahon2018linear} \citeyear{mcmahon2018linear}). We assume that a dose response that is greater or equal than 20 is critical for the organisms in the soil and aim to estimate $\mathbb{P}(\mathcal{R}(\para) \geq 20.0 | \data)$. As the prior and the likelihood are Gaussian and the measurements depend linearly on the property field~$\para$, there exists an analytical solution for the Gaussian posterior $p_{\para|\dataRV}(\para | \data)$. The PDF $p_{R|\dataRV}(r | \data)$ of the quantity of interest follows a generalized chi-square distribution (Fig.~\ref{EBM_fig:contamin_postR}). Even if this distribution does not have a simple closed-form expression, we can derive numerically the true probability of the considered rare event, which is $\mathbb{P}(\mathcal{R}(\para) \geq 20.0 | \data) =  1.76 \times 10^{-6}$. \\

We parameterize the bias potential $V_{\psi}(r)$ using RBFs with squared exponential kernels (Eq. \ref{EBM_RBF}). We assume a constant shape parameter $\kappa$, use evenly distributed centers $b_j$ from $L_B$ to $H_B$ and only optimize the weights $w_j$ during the optimization of $V_{\psi}(r)$. We choose $L_B = -80$ and $H_B = 120$ to ensure that the optimization of the bias potential is not limited by a too narrow range (high-mass regions of $p_{R|\dataRV}(r|\data)$ and $p_{\textup{ref}}(r)$ included). After testing different numbers of RBFs $B$ and shape parameters $\kappa$, we choose a parameterization with $B=500$ and $\kappa=1$. The chosen parametric form has to enable a reasonable approximation of the optimal bias potential. In practice, a poor choice becomes evident if the samples of $p_{V_\psi}(r)$ consistently diverge from the chosen $p_{\textup{ref}}(r)$. This discrepancy can be detected visually, or also through the empirical values of the $\textup{KL}(p_{\textup{ref}}||p_{V_\psi})$ (Eq. \ref{KLdiv}) or the $\textup{KSD}(p_{\textup{ref}}||p_{V_\psi})$ (Eq. \ref{KSD}). We also explore our non-parametric approach for modeling the bias potential (Section \ref{nonparV}), which involves employing a kernel density estimate to approximate the probability density function $p_V(r)$. To perform this estimation, we utilize the R function \textit{density} \citep{Rstats} with the `nrd' bandwidth configuration \citep{scott2015multivariate}. In practice, for numerical evaluations, we discretize the function by considering equidistant points within the range of -80 to 120, with a spacing of 0.1.  \\

We introduce a base scenario for the algorithmic setting. For $p_{\textup{ref}}(r)$, we use a Gaussian PDF $\mathcal{N}(20,7^2)$ centered on the critical threshold. We iterate for a maximum of 500 SGDM optimization steps employing a constant learning rate~$\gamma$ of 1.2 and a momentum weight~$\beta$ of 0.5 for the RBF parameterization (Eq. \ref{sgdm}). For the non-parametric form of the bias potential, we use the same momentum weight and a constant learning rate of 15. For the sampling of $R_i \sim p_{V_{\psi}}(\cdot)$ (Eq. \ref{EBM_loss_n}), we use MH steps employing Gaussian proposals with a step width chosen by initial testing to obtain an acceptance rate close to 30 $\%$. To account for the fact that the posterior is much better defined for the cells where measurements have been made, we adjust the step size for these specific cells. We utilize the KSD stopping criterion with a significance level of $\alpha = 95 \%$ and a bootstrap parameter $a_{BS}=0.4$ (Section~\ref{EBM_stop}). Following the recommendations of \citet{chwialkowski2016kernel}, we use $n=125$ final samples with a thinning factor of $th=10$ and a burn-in of 100 steps. \\

\begin{figure}	
	\begin{subfigure}[b]{0.48\textwidth}
         \centering
         \includegraphics[width=8cm]{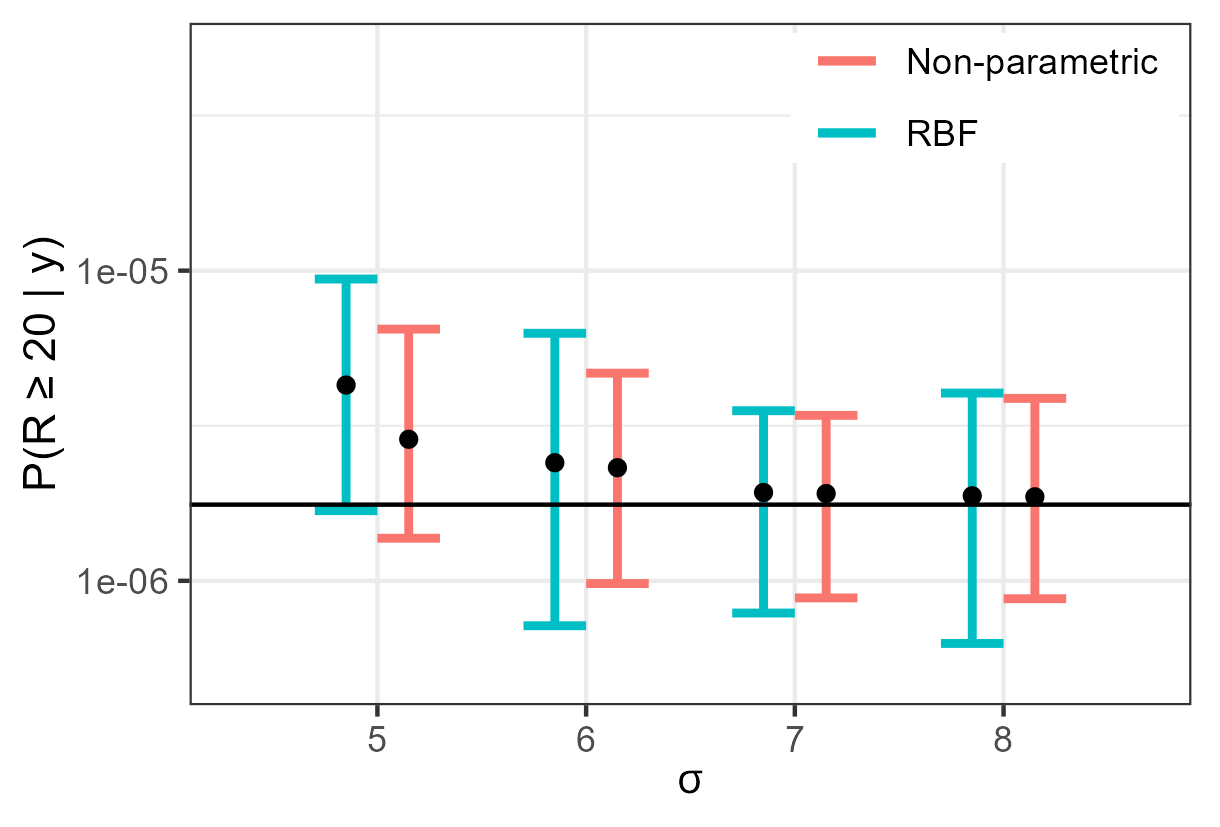}
         \caption{}
         \label{EBM_fig:EBM_p1}
     \end{subfigure} 
     \begin{subfigure}[b]{0.48\textwidth}
         \centering
         \includegraphics[width=8cm]{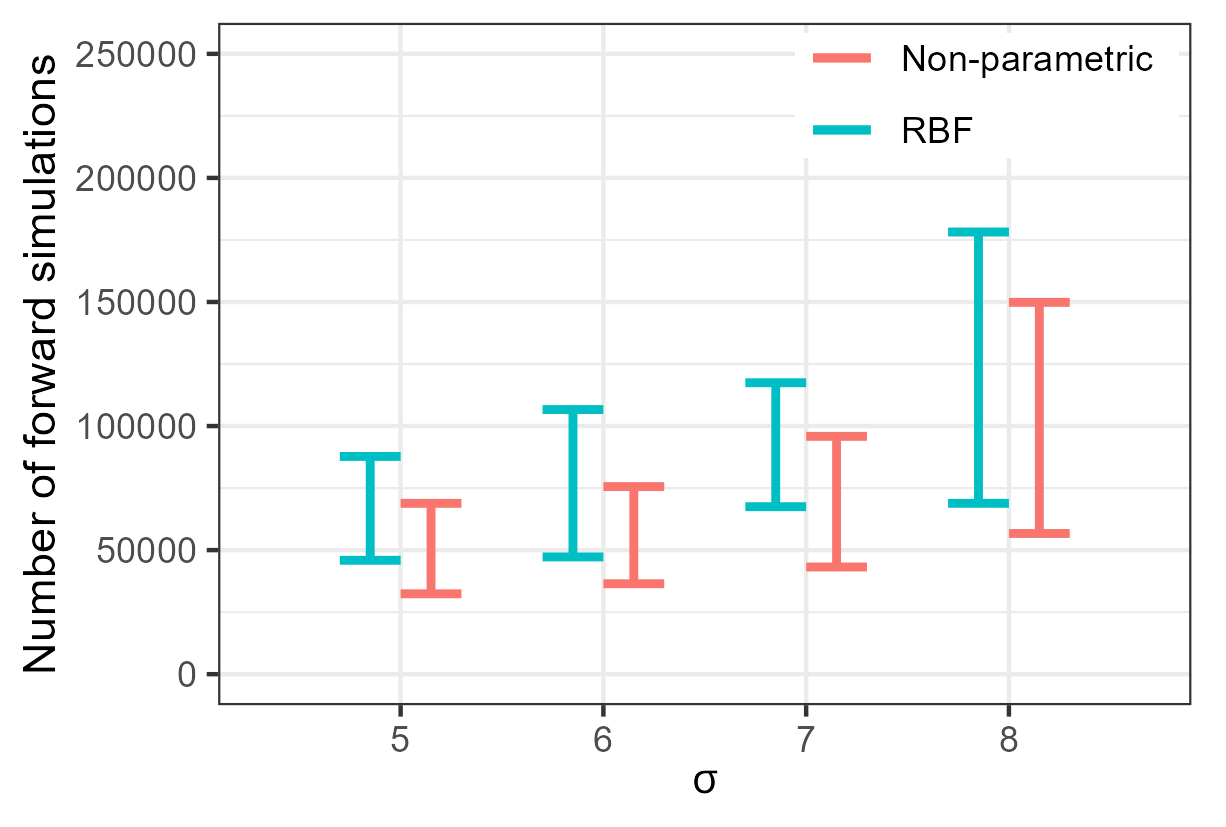}
         \caption{}
         \label{EBM_fig:EBM_p2}
     \end{subfigure}
		\caption{Analytical contamination example. (a) Range of rare event probability estimates obtained from 50 EBM runs using different standard deviations in defining $p_{\textup{ref}}(r)$, as well as parametric and non-parametric forms of the bias potential, (b) the corresponding minimum and maximum number of forward simulations. The horizontal line in (a) indicates the true rare event probability and the dots the mean of the 50 estimates. }
		\label{EBM_fig:EBM_p}				
\end{figure}

First, we examine various choices of $p_{\textup{ref}}(r)$ based on Gaussian distributions $\mathcal{N}(20,\sigma^2)$, where the standard deviation $\sigma$ ranges from 5 to 8. Figure~\ref{EBM_fig:EBM_p1} depicts the range of rare event probability estimates obtained from 50 runs using the different choices of $p_{\textup{ref}}(r)$ and both the RBF and the non-parametric estimation of the bias potential. For both forms, the stability and accuracy of the method becomes apparent from a standard deviation of 7 and above. The black points in Figure~\ref{EBM_fig:EBM_p1} depict the mean probability values of the 50 runs. We observe a positive bias that diminishes as the standard deviation of $p_{\textup{ref}}(r)$ increases. This bias in the estimate is more pronounced for the RBF bias potential. This arises from initializing $V(r)$ as a constant function. Throughout the optimization process, the weights of the RBF $V(r)$ are only modified up to the point where samples of $p_{\textup{ref}}(r)$ exist. For $r$ values exceeding a high quantile of $p_{\textup{ref}}(r)$, the resulting optimized bias potential is biased upwards. Consequently, $F(r) = - \log(p_{\textup{ref}}(r)) - V_{\textup{opt}}(r)$ is underestimated in that region. This causes $p_{R|\dataRV}(r|\data)$ and $\mathbb{P}(\mathcal{R}(\para) \geq T | \data)$ to be overestimated. This effect is less pronounced with the non-parametric bias potential, as the stochastic gradient is based on the analytical PDF $p_{\textup{ref}}(r)$. Figure~\ref{EBM_fig:EBM_p2} illustrates the minimum and maximum number of forward simulations used in the 50 runs. Despite the stability of the probability estimate's uncertainty from a standard deviation of 7, the number of required optimization steps continues to rise as the standard deviation increases. The RBF parameterization demands a greater number of simulations compared to the non-parametric approach and yields estimates with a higher variance. \\

\begin{figure}	
\centering
	\begin{subfigure}[b]{.32\textwidth}
         \centering
         \includegraphics[height=4.2cm]{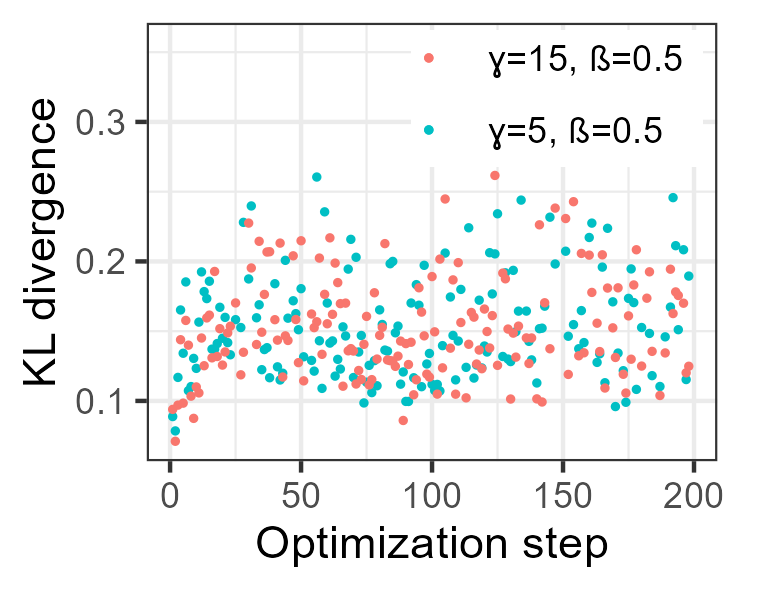}
         \caption{}
         \label{EBM_fig:EBM_LR11}
     \end{subfigure}
     \begin{subfigure}[b]{.32\textwidth}
         \centering
         \includegraphics[height=4.2cm]{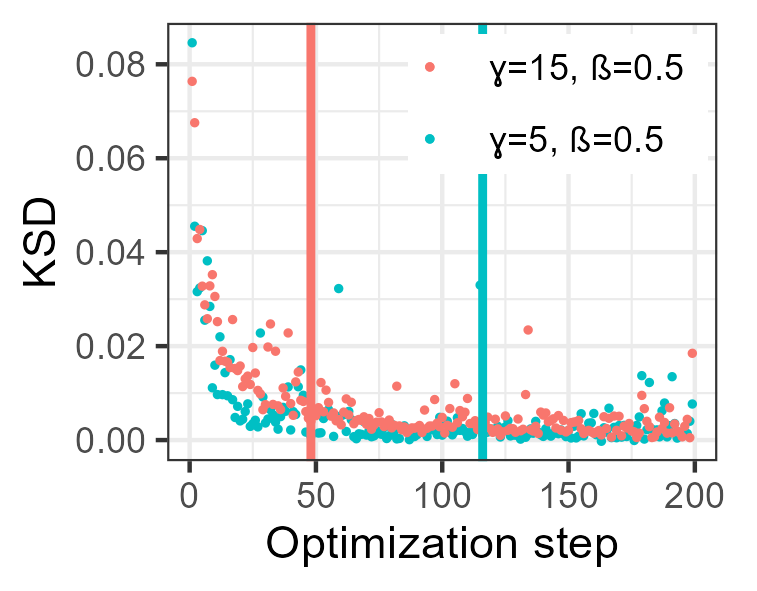}
         \caption{}
         \label{EBM_fig:EBM_LR12}
     \end{subfigure} \linebreak
     \begin{subfigure}[b]{.32\textwidth}
         \centering
         \includegraphics[height=4.2cm]{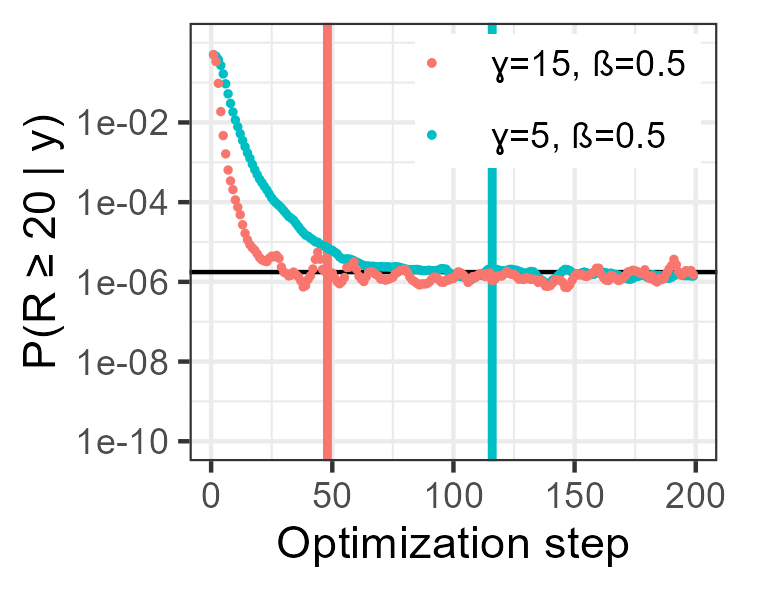}
         \caption{}
         \label{EBM_fig:EBM_LR13}
     \end{subfigure}
     \begin{subfigure}[b]{.32\textwidth}
         \centering
         \includegraphics[height=4.2cm]{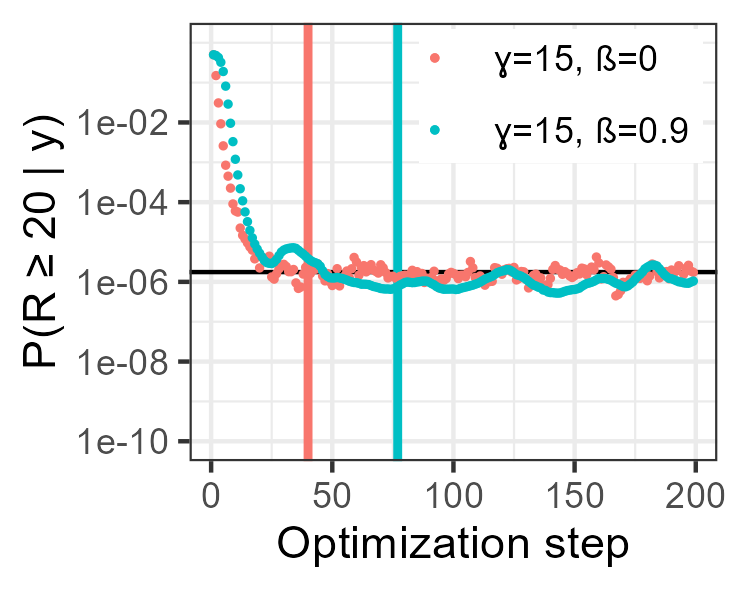}
         \caption{}
         \label{EBM_fig:EBM_LR14}
     \end{subfigure}
		\caption{Analytical contamination example. Example trajectories for the evolving (a) Kullback--Leibler divergence, (b) KSD and (c, d) rare event probability estimate using different learning rates $\gamma$ and momentum weights $\beta$. The horizontal lines in (c, d) indicate the true rare event probability and the vertical lines in (b, c, d) the step at which the KSD criteria would have halted the optimization process. }
		\label{EBM_fig:EBM_LR}				
\end{figure}

We now focus on the optimization scheme while maintaining $p_{\textup{ref}}(r)=\mathcal{N}(20,7^2)$ and the non-parametric form of the bias potential. Figures~\ref{EBM_fig:EBM_LR11}~-~\ref{EBM_fig:EBM_LR14} show example trajectories of the Kullback--Leibler divergence, KSD values and rare event probability estimates for different learning rates $\gamma$ and momentum weights $\beta$. We perform 200 optimization steps without any stopping criterion and highlight the step at which the KSD criteria would have halted the optimization process (indicated by vertical lines). We emphasize that while the Kullback--Leibler divergence exhibits a great amount of scatter, the KSD values demonstrate a much clearer convergence (Figs. \ref{EBM_fig:EBM_LR11} and \ref{EBM_fig:EBM_LR12}). The mean, root-mean-square-error (RMSE) and coefficient of variation (COV) of the estimates using a learning rate $\gamma=15$ and momentum weight $\beta=0.5$ are summarized in Table~\ref{EBM_tab:analytical}. It is observed that whether a stopping criterion is applied or not, the accuracy of the estimate is similar. However, the computational budget can be reduced by almost a factor of 4 when considering 200 fixed optimization steps. Increasing the learning rate comes with a trade-off: while it reduces the computational requirement by reducing the number of optimization steps needed, there is a corresponding loss in accuracy (Fig. \ref{EBM_fig:EBM_LR13}). The momentum weight plays a crucial role in mitigating the impact of stochastic gradient fluctuations. However, when paired with a too high learning rate, there is a risk of overshooting, necessitating careful adjustment of the learning rate for optimal performance (Fig. \ref{EBM_fig:EBM_LR14}, further discussed in Section \ref{EBM_ebm_loadcapa}).  \\

We compare the results of the EBM method with subset sampling (Section~\ref{EBM_MH_SS}). To perform subset sampling, we establish a fixed burn-in of 100 steps during the initial posterior estimation process. Additionally, we utilize a thinning factor of 500 for the MH samples to promote an initial sample representation that is free from correlation. To prevent bias from an adaptive sequence of thresholds and avoid re-running the algorithm, we use a fixed sequence of thresholds given by a logarithmic function increasing from 5.0 to $T$ within $nr_{t}$ steps. Then we assume the same total computational budget as for the EBM method (including the posterior sampling) and choose the number of samples per subset such as the number of MH steps per subset based on the lowest RMSE using 50 test runs. Here, we choose 60 samples per subset that are each propagated using 20 MH steps. The statistics of the resulting rare event probability estimates are summarized in Table~\ref{EBM_tab:analytical}. Both the lowest RMSE and COV values obtained with subset sampling are about 60 $\%$ higher than the ones of our EBM approach. \\

\begin{table}[]
\caption{Comparison of the EBM method and subset sampling for the analytical contamination example. Mean refers to the mean of the 50 estimates of $\mathbb{P}(\mathcal{R}(\para) \geq T | \data)$, RSME to their root-mean-square-error and COV to the coefficient of variation. The budget noted for EBM corresponds to the mean budget.  }
    \label{EBM_tab:analytical}
    \centering
    \def\arraystretch{1.5}
    \begin{tabular}{c c c c c}
	Method & Budget &  Mean & RMSE & COV   \\
	\hline
    EBM (Stopping criterion) & 72 k& $1.74 \times 10^{-6}$ & $0.62\times 10^{-6}$& 0.36 \\
    EBM (200 iterations) & 270 k& $1.79 \times 10^{-6}$ & $0.62 \times 10^{-6}$& 0.35 \\
    Subset sampling & 72 k & $1.73 \times 10^{-6}$ & $1.00 \times 10^{-6}$ & 0.58 \\
\end{tabular}
\end{table}

\subsection{Four-branch function}
\label{EBM_ebm_branch}
\begin{figure}	
\centering
\begin{subfigure}[b]{.3\textwidth}
         \includegraphics[height=5cm]{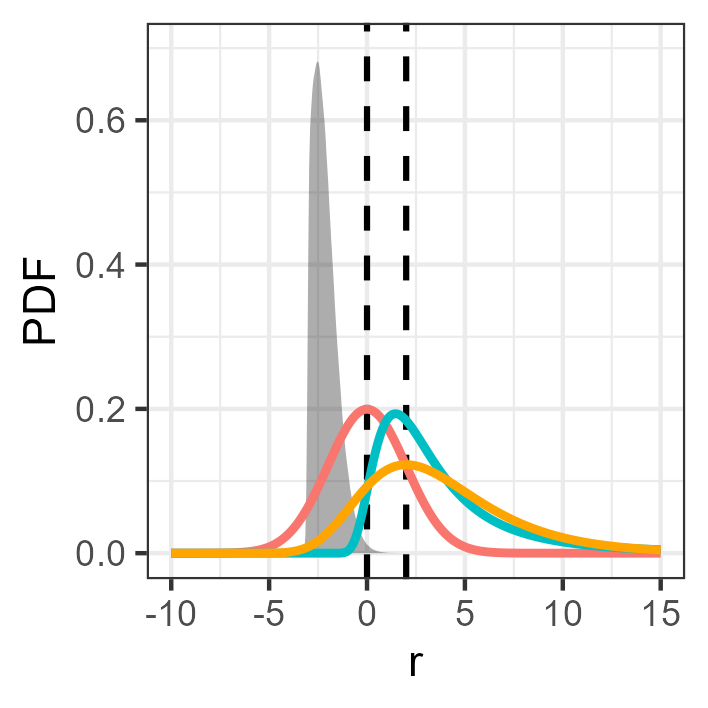}
         \caption{}
         \label{EBM_fig:EBM_branch_N1}
     \end{subfigure}
    \begin{subfigure}[b]{.22\textwidth}
         \includegraphics[height=4cm]{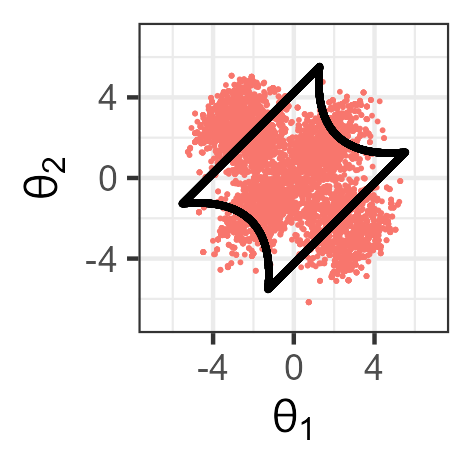}
         \hspace{.5cm}
         \caption{}
         \label{EBM_fig:EBM_branch_N2}
     \end{subfigure}
     \begin{subfigure}[b]{.22\textwidth}
         \includegraphics[height=4cm]{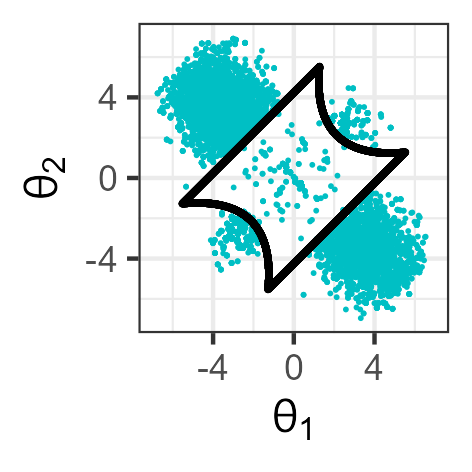}
         \hspace{.5cm}
         \caption{}
         \label{EBM_fig:EBM_branch_N3}
     \end{subfigure} 
     \begin{subfigure}[b]{.22\textwidth}
         \includegraphics[height=4cm]{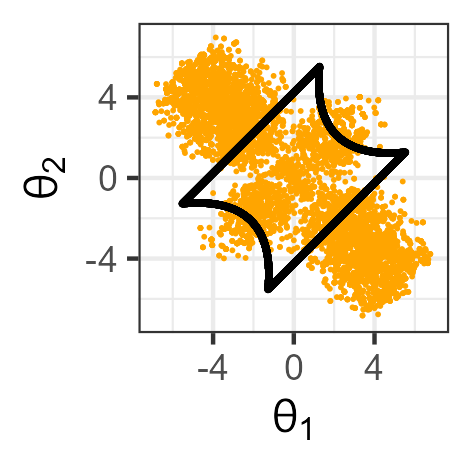}
         \hspace{.5cm}
         \caption{}
         \label{EBM_fig:EBM_branch_G3}
     \end{subfigure}
		\caption{Impact of $p_{\textup{ref}}(r)$ for the four-branch function example. (a) Distribution of the quantity of interest $R$ (filled) with two considered thresholds (dashed vertical), along with the PDFs $\mathcal{N}(0,2^2)$ (red), $\mathrm{GEV}(2,2,0.33)$ (blue) and $\mathrm{GEV}(2,3,0)$ (orange). A total of 10,000 samples of the resulting trained $p_{V}(\theta_1, \theta_2)$ obtained with (b) $\mathcal{N}(0,2^2)$, (c) $\mathrm{GEV}(2,2,0.33)$, and (d) $\mathrm{GEV}(2,3,0)$. The black outlines in (b, c, d) depict the limit states $\para$ with $\mathcal{R}(\para) = T^*$. }
		\label{EBM_fig:EBM_branch_p}				
\end{figure}

The four-branch function represents a widely used benchmark in structural reliability analysis, characterizing the failure of a system comprising four distinct component limit states (e.g., \citeauthor{schobi2017rare} \citeyear{schobi2017rare}). It is used here to demonstrate the applicability of our EBM method outside of an inversion setting. The four-branch function is defined as,
\begin{equation}
    \mathcal{R} (\para) = \min \left\{
    \begin{array}{l}
    3 + 0.1(\theta_1 - \theta_2)^2 - \frac{\theta_1+\theta_2}{\sqrt{2}} \\
    3 + 0.1(\theta_1 - \theta_2)^2 + \frac{\theta_1+\theta_2}{\sqrt{2}} \\
    (\theta_1-\theta_2) + \frac{6}{\sqrt{2}} \\
    (\theta_2-\theta_1) + \frac{6}{\sqrt{2}} \\
    \end{array}
    \right\}, 
\end{equation}
with input variables $\para = (\theta_1, \theta_2)$, which are modeled by two independent standard normally-distributed variables. The traditional rare probability of interest is $\mathbb{P}(\mathcal{R}(\para) \leq T^*) = \mathbb{P}(- \mathcal{R}(\para) \geq T^*)$, with $T^*=0$. Given that with probability $4.46 \times 10^{-3}$ (according to a MC estimation with $10^8$ samples in \citeauthor{schobi2017rare} \citeyear{schobi2017rare}) this is not a particularly rare event, we also direct our focus towards $\mathbb{P}(- \mathcal{R}(\para) \geq T^{**})$ with $T^{**} = 2$. The distribution of the quantity of interest $R = - \mathcal{R}(\para)$ is depicted in Figure~\ref{EBM_fig:branch_postR} and the limit states $\para$ with $\mathcal{R}(\para) = T^*$ are depicted in Figure~\ref{EBM_fig:EBM_branch_p}b-d as black outlines.\\

We use the non-parametric form of the bias potential with the same Gaussian kernel density estimate as in the previous section. As evaluation points, we employ equally-spaced intervals between -10 and 100 with a spacing of 0.1. Our objective is to investigate the performance of the EBM method under a restricted computational budget of 10,000 evaluations. To perform the sampling of $R_i \sim p_{V_{\psi}}(\cdot)$, we use MH steps with Gaussian proposals aiming for an acceptance rate close to 30 $\%$. We test a configuration using a stopping criterion employing a significance level of $\alpha=0.99$ and $a_{BS}=0.5$. With an appropriate thinning factor (4) and burn-in period (5 steps), this results in 100 samples obtained during 405 MH iterations. We run the EBM process with a constant learning rate of $\gamma=6.5$ and momentum weight $\beta=0.5$. \\

The optimized bias potential $V_{\textup{opt}}(r)$ takes on different forms depending on the chosen PDF $p_{\textup{ref}}(r)$, impacting not only $p_{V_{\textup{opt}}}(r)$, but also $p_{V_{\textup{opt}}}(\para)$ (Eq. \ref{EBM_pv_theta}), which has to be explored by MH steps to obtain the samples $R_i \sim p_{V_{\psi}}(\cdot)$. This is particularly interesting in this test example, as the failure event is connected to four distinct regions of the parameter space. Figure~\ref{EBM_fig:EBM_branch_p} shows samples of the resulting joint PDFs $p_{V_{\textup{opt}}}(\theta_1, \theta_2)$ for different choices of $p_{\textup{ref}}(r)$ together with the limit states $\para$ with $\mathcal{R}(\para) = T^*$ (black outlines). The corresponding means and COVs for the rare event probability estimates are shown in Table~\ref{EBM_tab:branch}. We start with a Gaussian distribution centred on the first threshold of interest, $\mathcal{N}(0,2^2)$. The resulting joint PDF $p_{V_{\textup{opt}}}(\theta_1, \theta_2)$ (Fig.~\ref{EBM_fig:EBM_branch_N2}) appears to be relatively compact and only explores a small part of the area outside the limit states. This leads to a high mean estimate (Table~\ref{EBM_tab:branch}). Merely increasing the standard deviation means that $p_{\textup{ref}}(r)$ also puts high mass on the left tails of the distribution of $R$, a wasteful allocation of resources. \\

As an alternative, we consider an asymmetric and heavy-tailed Generalized Extreme Value (GEV) distribution as $p_{\textup{ref}}(r)$. We start with $\mathrm{GEV}(2,2,0.33)$ using a location parameter of 2, a scale parameter of 2, and a shape parameter of 0.33. This shift results in a PDF $p_{\textup{ref}}(r)$ that has minimal mass in the high-probability region of $R$ (Fig.~\ref{EBM_fig:EBM_branch_N1}). The resulting joint PDF $p_{V_{\textup{opt}}}(\theta_1, \theta_2)$ (Fig.~\ref{EBM_fig:EBM_branch_N3}) displays an increased level of exploration outside the limit states but partly omits the central region that links the four branches. This creates challenges when transitioning between modes during the MH sampling process, leading to biased probability estimates with high COVs (Table~\ref{EBM_tab:branch}). It also required a slightly increased computational budget of 12,000 evaluations. These findings suggest that a well-chosen $p_{\textup{ref}}(r)$ should not only have heavy right tails, but also include the high probability region of $R$ while excluding the left tails of its distribution. Therefore, we change to a type I extreme value distribution $\mathrm{GEV}(2,3,0)$ (Fig.~\ref{EBM_fig:EBM_branch_N1}), which expands the joint PDF $p_{V_{\textup{opt}}}(\theta_1, \theta_2)$ while the central region is still incorporated (Fig.~\ref{EBM_fig:EBM_branch_G3}). Consequently, our results yield a mean probability estimate for $T^*$ that aligns with the one in \citet{schobi2017rare}. Moreover, for $T^{**}$, we observe a significantly reduced mean compared to the alternative $p_{\textup{ref}}(r)$, and for both thresholds, the COVs are at their lowest with the latter choice.\\

To ensure a connected $p_{V_{\textup{opt}}}(\theta_1, \theta_2)$ and enhanced exploration of the failure regions, we consider $\mathrm{GEV}(2,3,0)$ to be the best choice. In evaluating subset sampling for comparison purposes, we employ the average computational budget of the EBM approach. A satisfactory performance was achieved with 80 samples per subset and a propagation using 5 MH steps. For both thresholds, we obtain comparable mean estimates with both methods, yet higher COVs with subset sampling compared to EBM with $\mathrm{GEV}(2,3,0)$. In fact, the COV for the second threshold $T^{**}$ is about 50 $\%$ higher. \\

\begin{table}[]
\caption{Comparison of the EBM method and subset sampling for the four-branch example. Mean refers to the mean of the 50 estimates of $\mathbb{P}(-\mathcal{R}(\para) \geq T)$ for the two thresholds $T^*$ and $T^{**}$ and COV to the coefficient of variation. The budget noted for EBM is equal to the mean budget. The Monte Carlo results for $T^*$ are copied from \citet{schobi2017rare}. }
    \label{EBM_tab:branch}
    \centering
    \def\arraystretch{1.5}
    \begin{tabular}{c c c c c c c}
	Method & $p_{\textup{ref}}(r)$ & Budget &  Mean $T^*$ & COV $T^*$ & Mean $T^{**}$ & COV $T^{**}$ \\
 \hline
 MC & - & $10^5$ k & $4.46 \times 10^{-3}$ & 0.0015& - & - \\
     \hline
     EBM & $\mathcal{N}(0,2^2)$ & 5 k & $5.79\times 10^{-3}$ & 0.47 & $45.69 \times 10^{-5}$ & 0.63  \\
     EBM & $\mathrm{GEV}(2,2,0.33)$ & 12
 k & $82.45 \times 10^{-3}$ & 1.63 & $31.29 \times 10^{-5}$ & 1.09  \\
    EBM & $\mathrm{GEV}(2,3,0)$ & 10 k & $4.97\times 10^{-3}$ & 0.31 & $1.56 \times 10^{-5}$ & 0.53  \\
    \hline
   Subset sampling & - & 10 k &  $4.91 \times 10^{-3}$ & 0.36 & $1.36\times 10^{-5}$ & 0.80 \\
    \hline
\end{tabular}
\end{table}

\subsection{Load capacity example}
\label{EBM_ebm_loadcapa}
As a third test case, we consider the load capacity example of \citet{straub2016bayesian}, enabling comparison of our EBM approach with their BUS method (Section~\ref{EBM_BUS}). The problem under consideration is a basic reliability problem concerned with the failure of an engineering system when the load $\para^L$ exceeds the capacity $\para^C$. The corresponding rare event probability of interest is,
\begin{equation}
    \mathbb{P}(\mathcal{R}(\para) \geq 0) = \mathbb{P}(\para^L - \para^C \geq 0),
\end{equation}
with $\para = (\para^L,\para^C)$. For the load $\para^L$, we assume a prior Gumbel distribution with mean 2 and standard deviation 1 and the capacity $\para^C$ is lognormal distributed with mean 12 and standard deviation 2; the variables are assumed to be independent. To investigate the performance for a problem with increasing number of model parameters, \citet{straub2016bayesian} express the capacity as a product of component capacities $\para^C_i$ ($1 \leq i \leq n_C$). The component capacities $\para^C_i$ follow a lognormal distribution such that $\para^C$ still has mean 12 and standard deviation 2. \\

In this inversion setting, individual measurements $y_i$ of the component capacities $\para^C_i$ are considered. To ensure analytical solutions, multiplicative lognormal measurement errors,
\begin{equation}
    p_{\dataRV|\para}(\data | \para) = p(\data | \para^C_1, \para^C_2,..., \para^C_{n_C}) \propto \exp \left(- \frac{1}{2} \sum_{i=1}^{n_C} \left( \frac{\log y_i - \log \para^C_i}{\sigma_y} \right)^2 \right),
\end{equation}
with $\sigma_y = 0.05$ and $y_i = 8^{1/n_C}$ are employed. Following \citet{straub2016bayesian}, we consider $n_C=10$ and $n_C=100$ with the resulting failure probabilities $\mathbb{P}(\para^L - \para^C \geq 0 | \data)$ being $6.8 \times 10^{-5}$ and $2.1 \times 10^{-5}$, respectively. The posterior distribution of $R = \mathcal{R}(\para) = \para^L - \para^C$ for $n_C=10$ with the critical threshold of $T=0$ is depicted in Figure~\ref{EBM_fig:load_postR}. \\

The results of the BUS approach with subset sampling are reproduced in Table~\ref{EBM_tab:loadcapa}. As \citet{straub2016bayesian}, we transform the reliability problem to the standard normal space by marginal transformations of the load and capacity variables. To enable a direct comparison of the EBM with the BUS approach, we employ the same computational budgets. In order to accommodate a consistent and constrained computational budget in the context of EBM, we use a fixed number of optimizations steps (no stopping criterion). Following our analysis in the previous section, we employ as $p_{\textup{ref}}(r)$ a type I extreme value distribution $\mathrm{GEV}(0,7,0)$. To make the method dimensionally robust, we use prior-preserving preconditioned Crank-Nicolson proposals (pCN; e.g. \citeauthor{cotter2013} \citeyear{cotter2013}) within the sampling of $R_i \sim p_{V_{\psi}}(\cdot)$. Then, we use a burn-in of 10 steps, a thinning factor of $th=6$ and $a_{BS}=0.5$. As we do not rely on a stopping criterion, we further divide the number of final samples by a factor of two \citep{chwialkowski2016kernel}, leading to $n = 50$ obtained by 310 MH steps. Then we fix the number of optimization steps in accordance with the computational budget (resulting in 25 and 27 steps for the two values of $n_C$). \\

\begin{figure}	
\begin{subfigure}[b]{.48\textwidth}
         \centering
         \includegraphics[height=5cm]{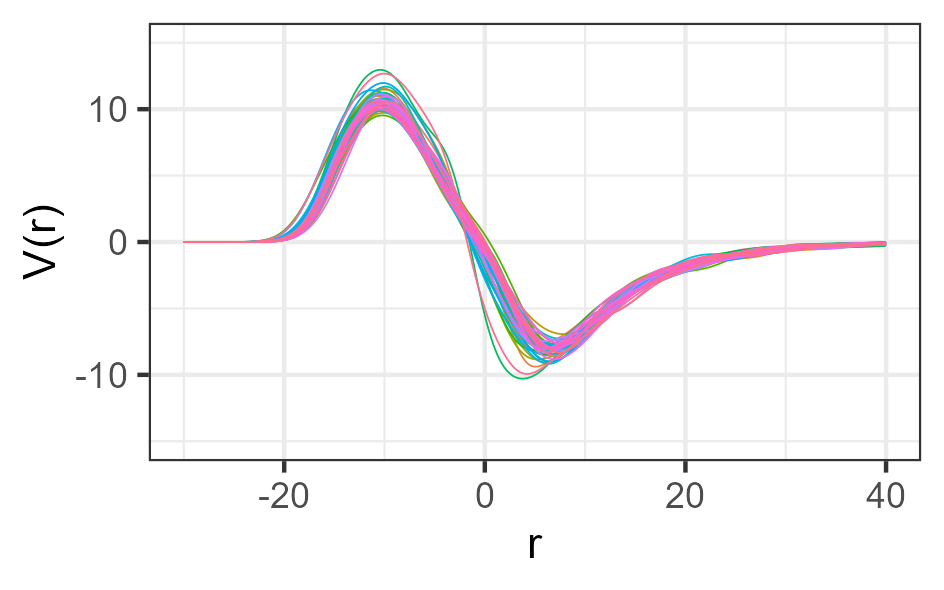}
         \caption{}
         \label{EBM_fig:EBM_load1}
     \end{subfigure}
    \begin{subfigure}[b]{.48\textwidth}
         \centering
         \includegraphics[height=5cm]{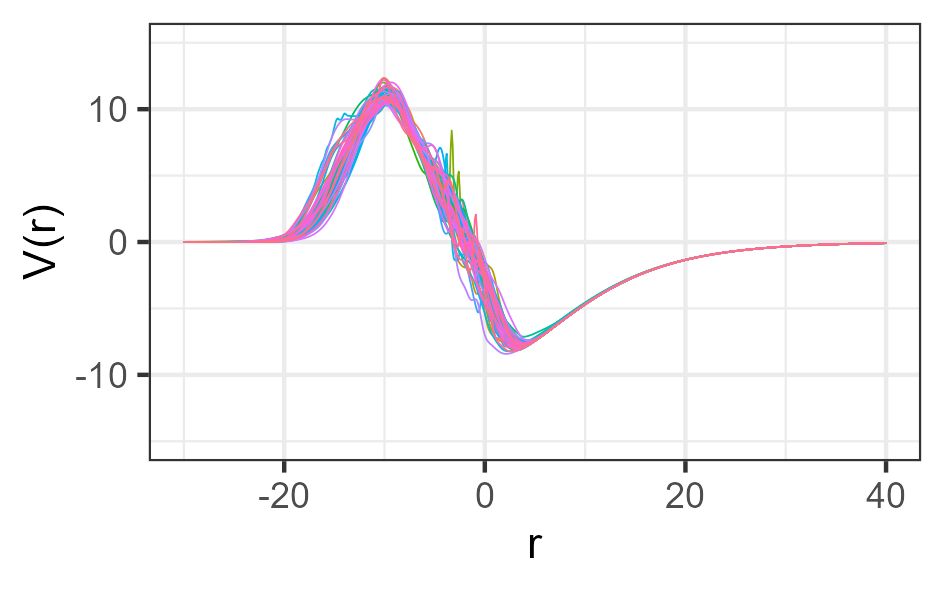}
         \caption{}
         \label{EBM_fig:EBM_load2}
     \end{subfigure}
		\caption{Twenty optimized bias potentials $V(r)$ for the load capacity example ($n_C=100$) using (a) the RBF parameterization and (b) the non-parametric form. }
		\label{EBM_fig:EBM_load}				
\end{figure}

We test both a parametric and non-parametric form for the bias potential. For the parametric approach, we employ RBFs with squared exponential kernels (Eq. \ref{EBM_RBF}), using $L_B = -100$ and $H_B = 100$, $B=500$ and $\kappa=0.5$. We run the EBM process with an exponentially decreasing learning rate initiated as $\gamma=0.5$ (factor -0.005) and momentum weight $\beta=0.5$. For the non-parametric form, we use the same Gaussian kernel density estimate of $p_V(r)$ as in the previous sections and evaluation points between -100 and 100 with a spacing of 0.1. Here we employ an exponentially decreasing learning rate initiated as $\gamma=19$ (factor -0.005) and momentum weight $\beta=0.95$. The means and 95 $\%$ confidence intervals of 50 runs are shown in Table \ref{EBM_tab:loadcapa} and the corresponding estimates of $V(r)$ are depicted in Figure \ref{EBM_fig:EBM_load1} for the RBFs and in Figure \ref{EBM_fig:EBM_load2} for the non-parametric form. With a non-parametric $V(r)$, we sometimes obtain bias potentials characterized by minor-scale fluctuations, in contrast to the smoother bias potentials generated by RBFs, which exhibit more pronounced large-scale fluctuations. The small-scale fluctuations arise due to the constraints of our limited computational resources, compelling us to base the kernel density estimates on a relatively small set of samples, which may not be adequately thinned. To mitigate this issue and promote a smoother estimation, we have increased the momentum weight to $\beta=0.95$. However, this adjustment also introduces challenges in fine-tuning the learning process, as it tends to cause overshooting during optimization. Due to the changed PDF $p_{R|\dataRV}(r|\data)$ and computational budget for $n_C=100$, we have to adapt the learning rate schedule for an optimal performance and use $\gamma=22$ and a decay factor of -0.003. On the other hand, the RBF parameterization with lower momentum weight is less sensitive to the choice of the learning rate schedule and we can use the same for both scenarios. For both $n_C$, both presented configurations of the EBM approach lead to a reduction of the 95 $\%$ confidence interval's length compared to the BUS approach. While the non-parametric form, when coupled with a higher momentum weight, yields noticeably increased accuracy, it also becomes significantly more sensitive to the learning rate schedule. \\

\begin{table}[]
\caption{Table summarizing the comparison of EBM and BUS for the load capacity example. The 95~$\%$ confidence intervals of the BUS method with subset sampling are copied from \citet{straub2016bayesian}, the mean value is not provided.  }
    \label{EBM_tab:loadcapa}
    \centering
    \def\arraystretch{1.5}
    \begin{tabular}{c c c c c}
	Case & Method & Budget & Mean & \makecell{95 $\%$ confidence interval }   \\
	\hline
        $n_C = 10$ & Analytical & - & $6.8 \times 10^{-5}$ &- \\
        & \makecell{EBM (non-par.)} & 7.7 k & $7.0 \times 10^{-5}$ & $[2.7,12.0] \times 10^{-5}$ \\
        & EBM (RBF) & 7.7 k & $7.1 \times 10^{-5}$  & $[0.9,19.5] \times 10^{-5} $\\
        & BUS & 7.7 k & - & $[2.1,35.0] \times 10^{-5}$\\
        \hline
        
        $n_C = 100$ & Analytical & - & $2.1 \times 10^{-5}$ & - \\
        & \makecell{EBM (non-par.)} & 8.4 k & $2.8 \times 10^{-5}$ & $[1.1,5.0] \times 10^{-5}$ \\
        & EBM (RBF) & 8.4 k & $4.0 \times 10^{-5}$  & $[0.3,10.0] \times 10^{-5} $\\
        & BUS & 8.4 k & - & $[0.2,12.0] \times 10^{-5}$ \\
\end{tabular}
\end{table} 

\section{Discussion}
\label{EBM_ebm_disc}
This paper presents a novel energy-based model (EBM) approach for estimating probabilities of rare events that is applicable not only for conventional rare event estimation but also in inversion settings. The approach centers on representing the distribution of the quantity of interest as an energy density function. The estimation of the related free energy is achieved by optimizing a bias potential, aligning it with a predefined distribution $p_{\textup{ref}}(r)$ that has most mass on the region of interest.\\

When parameterizing the bias potential $V_{\psi}(r)$, it is crucial to allow for an appropriate amount of flexibility. We recommend to compare results obtained by a chosen parameterization with those obtained with a more flexible parameterization. An alternative approach is to use a non-parametric form for the bias potential $V(r)$. Depending on the test case and the computational budget, we illustrate the effectiveness of the non-parametric bias potential in alleviating bias issues and reducing computational needs (Fig. \ref{EBM_fig:EBM_p}). However, the non-parametric approach relies heavily on the quality of the kernel density estimate used during the optimization. In cases where computational resources are limited, and when dealing with correlated samples, it is possible to obtain bias potentials that exhibit fine-scale fluctuations resulting from individual optimization steps (Fig. \ref{EBM_fig:EBM_load2}). A parameterization, on the other hand, does not face this problem as it can be chosen such that it enforces a smooth bias potential (Fig. \ref{EBM_fig:EBM_load1}). However, in selecting a parameterization, we confront the task of determining the appropriate form, a choice that can substantially influence the results. When manually fine-tuning the bandwidth of the kernel density estimate within the non-parametric approach, we also retain the capability to shape the potential's form, but this comes at the cost of introducing additional configuration choices.  \\

When selecting the distribution $p_{\textup{ref}}(r)$, one needs to consider that too light right tails lead to a positive bias in the estimates of $\mathbb{P}(\mathcal{R}(\para) \geq T | \data)$. This effect is the most pronounced for the parametric approach, as the optimization relies on samples of $p_{\textup{ref}}(r)$ and $p_{V}(r)$. Beyond a certain quantile, the probability of obtaining samples is extremely small, leading to an overestimation of $V(r)$ in this region (under the assumption that the initial potential is constant everywhere). The same principle would hold for $\mathbb{P}(\mathcal{R}(\para) \leq T | \data)$ and the left tails of $p_{\textup{ref}}(r)$. Since our primary focus is on failure probability estimates, a slight positive bias in the probability value is not too problematic. Additionally, the non-parametric form reduces this bias as the optimization relies on PDFs enabling an integration of the tails. The choice of $p_{\textup{ref}}(r)$ also determines the form of the PDF $p_{V}(\para)$ (Eq. \ref{EBM_pv_theta}), which needs to be explored by MH in each optimization step. In the second test example utilizing the four-branch function (Section~\ref{EBM_ebm_branch}), we showcase the benefit of using an asymmetric Generalized Extreme Value (GEV) distribution as $p_{\textup{ref}}(r)$ (Fig.~\ref{EBM_fig:EBM_branch_p}). The form of the GEV distribution enables a $p_{\textup{ref}}(r)$ with heavy right tails that includes the high probability region of $R$. For this example with distinct regions contributing to the rare event, the incorporation of the region of high probability of the quantity of interest within $p_{\textup{ref}}(r)$ facilitates the MH algorithm's ability to shift between the modes, leading to a lower coefficient of variation in the probability estimates. \\

We optimize the bias potential using stochastic gradient descent with momentum. Regarding the learning rate, a smaller value tends to lead to more stable estimates but also to a larger number of optimization steps (Fig.~\ref{EBM_fig:EBM_LR}). Conversely, higher learning rates result in increased fluctuations but require fewer steps for convergence. The momentum has the advantage of dampening oscillations in the convergence process and making it more robust to noisy gradients, however, it requires careful tuning of the learning rate and momentum weight, to achieve optimal performance. Besides stochastic gradient descent with momentum, there exist numerous alternative implementations for the optimization. For instance, for the last test case requiring fine tuning of the optimization schedule (Section \ref{EBM_ebm_loadcapa}), exploring more advanced optimization algorithms like Adam \citep{kingma2014adam} would be interesting. We combine the optimization with a stopping criterion based on a goodness-of-fit test employing the Kernelized Stein discrepancy. This is done as the Kullback--Leibler divergence used in the optimization of the bias potential exhibits a great amount of scatter compared with KSD that demonstrates a much clearer convergence. This characteristic of the KSD makes it a valuable diagnostic tool in determining the appropriate moment to stop the optimization process. This achieves two key benefits: First, it helps stabilize the rare event probability estimates by ensuring that the biased probability distribution $p_V(r)$ closely approximates the target probability distribution $p_{\textup{ref}}(r)$. Second, it reduces computational effort by allowing the optimization process to halt as soon as the approximation becomes sufficiently accurate (Fig.~\ref{EBM_fig:EBM_LR}). To account for the correlation in the samples, we use the specific bootstrap method proposed by \citet{chwialkowski2016kernel}. \\

We compare the performance of the EBM approach against alternative methods (Section~\ref{EBM_alternatives}) using three illustrative test examples. In the first contamination test case involving inversion (Section~\ref{EBM_ebm_analytical}), the EBM method outperforms subset sampling, reducing the root-mean-square-error (RMSE) and coefficient of variation (COV) by about 40 $\%$ (Table~\ref{EBM_tab:analytical}). In the second example involving the four-branch function (Section~\ref{EBM_ebm_branch}), subset sampling also exhibits poorer performance than the EBM approach, with the latter reducing the COV by up to 35 $\%$ compared to subset sampling. In the third example in Section~\ref{EBM_ebm_loadcapa}, we compare the EBM method's performance against the BUS approach with subset sampling in a simplistic engineering test case from \citet{straub2016bayesian}. For both numbers of considered capacity components, the EBM method is able to narrow down the 95 $\%$ confidence interval compared to the BUS approach (Table~\ref{EBM_tab:loadcapa}). \\

Regarding different approaches for sampling $R_i \sim p_{V_{\psi}}(\cdot)$, we experiment with various bootstrap parameters $a_{BS}$ (results not shown). We notice that there is an increase in the variance of the probability estimates with growing $a_{BS}$. However, the reduced values for small bootstrap parameters come at the expense of a higher number of forward simulations. Similar to the learning rate, there seems to be a trade-off between computational cost and accuracy. Besides testing other parameterizations for $V(r)$, distributions for $p_{\textup{ref}}(r)$ and optimization schemes for updating $V(r)$, one could also consider replacing the sampler for $p_V(\para)$. While the third test example considering different numbers of component capacities (Section \ref{EBM_ebm_loadcapa}) touches upon the methods' capability to handle an increasing dimension of the underlying space of $\para$, it would be interesting to investigate this in more depth. For high-dimensional, complex posterior distributions, efficiently and accurately sampling the current biased posterior of $\theta$ using an MH algorithm can be challenging. This might render the EBM method computationally demanding, as each optimization step would require thousands of MH steps. However, this could be counteracted, for instance, using elaborated proposal schemes, interacting MH chains or particle methods (e.g., \citeauthor{robert2018accelerating} \citeyear{robert2018accelerating}). In data rich settings with narrow posterior distributions, it is expected that $V(r)$ will vary over very large ranges. In such scenarios, it could be beneficial to know in advance the expected properties of $V(r)$ such as shape, smoothness class and range. This could potentially be achieved by using a simplistic free energy function estimate (e.g. histogram based) of some thousands of posterior samples giving together with $p_{\textup{ref}}(r)$ to obtain a first approximation of $V_{\mathrm{opt}}(r)$. \\

\section{Conclusion}
\label{EBM_ebm_conc}
This paper introduces an energy-based model (EBM) approach to estimate rare event probabilities. The approach is based on formulating the distribution of the quantity of interest as an energy density function with a corresponding free energy function. By optimizing a bias potential such that the corresponding energy density approaches a pre-defined PDF $p_{\textup{ref}}(r)$, the method estimates the free energy accurately in the region targeted by $p_{\textup{ref}}(r)$. The presented approach is applicable both for traditional rare event estimation and in the context of inversion settings when one is not interested in the posterior itself, but rather in the distribution of a quantity that depends on the posterior. When employed in such a setting, this formulation reduces the potentially high-dimensional problem of first estimating the posterior and subsequently the quantity of interest to the optimization of a one-dimensional function. The optimization of the bias potential involves minimizing the Kullback--Leibler divergence and a stopping criterion based on the Kernelized Stein Discrepancy is introduced to terminate the optimization process. The stopping criterion not only enhances the stability and accuracy of the rare event probability estimation but also optimizes computational resources by terminating the optimization process when the approximation is deemed satisfactory. A non-parametric form of the bias potential is introduced, which eliminates the need to make a parameterization choice while simultaneously enabling efficient and accurate probability estimates. For the three presented test cases, a properly configured EBM approach ensures precise estimations of rare event probabilities and outperforms the examined variants of subset sampling methods.\\

This work was supported by the Swiss National Science Foundation (project number: \href{http://p3.snf.ch/project-184574}{184574}). The code associated with this article is available on 
\href{https://github.com/LeaFrie/EBM_rareevents}{github.com/LeaFrie/EBM\_rareevents}. \\

\appendix

\section{Relation to maximum likelihood estimation}
\label{EBM_app:MLE}
We can reformulate the minimization of the Kullback-Leibler divergence as a maximum likelihood estimation problem. While we summarize the general concepts and implications of this relation, we refer to \citet{van2000asymptotic} for the theoretical details about maximum likelihood estimation. \\

Let $R$ be a random variable, whose distribution $P$ has a PDF $p_{V_{\psi}}(r)$ with a statistical model $\{p_{V_{\psi}}(\cdot): \psi \in \Psi \}$. Furthermore, $r_i$ for $1 \leq i \leq n$ are $n$ independently generated realizations of $R$. The log-likelihood function of the model is given by,
\begin{equation}
    l_n(\psi) = \log \prod_{i=1}^n p_{V_\psi}(r_i) = \sum_{i=1}^n \log p_{V_\psi}(r_i).
\end{equation}
A maximum likelihood estimator (MLE) is defined as $\hat{\psi}_{MLE} \in \Psi$ with $l_n(\hat{\psi}_{MLE}) = \underset{\psi \in \Psi}{\max } \hspace{0.2cm} l_n(\psi)$.\\

We define, 
\begin{equation}
    \ell(\psi) := \mathbb{E}_{\psi_0}(\log p_{V_{\psi}}(R)) = \int \log p_{V_{\psi}}(r) p_{V_{\psi_0}}(r) \mathrm{d} r.
\end{equation}
Under the assumption that the model is well specified with $p_{V_{\psi_0}}(r)$ being the PDF of $P$ and \linebreak $\mathbb{E}(|\log p_{V_{\psi}}(R)|)<\infty$, $\psi \mapsto \ell(\psi)$ is maximized at $\psi_0$. The normalized log-likelihood $\psi \mapsto \frac{1}{n} l_n(\psi)$ is a sample approximation of $\psi \mapsto \ell(\psi)$, and under some regularity constraints, $\frac{1}{n}\l_n(\psi) \overset{a.s.}{\longrightarrow} \mathbb{E}_{\psi_0} \large( \log p_{V_\psi}(R) \large)$ almost surely such that the MLE $\hat{\psi}_{MLE}$ is a consistent estimator for $\psi_0$. Furthermore, under some additional conditions, the MLE follows asymptotic normality,
\begin{equation}
    \sqrt{n}(\hat{\psi}_{MLE}  -\psi_0) \overset{d}{\longrightarrow} \mathcal{N}(0,I(\psi_0)^{-1}), \quad \text{for  } n \rightarrow \infty,
\end{equation}
with the Fisher information matrix $I(\psi_0) = \mathbb{E}_{\psi_0} \Big( \nabla_\psi\log p_{V_{\psi}}(R) \nabla_\psi\log p_{V_{\psi}}(R)^T \Big)$. \\

In our setting, we assume that $p_{\textup{ref}}(r) = p_{V_{\psi_0}}(r)$ for $\psi_0 \in \Psi$. Furthermore, the Kullback-Leibler divergence between two distributions with PDFs $p_{V_\psi}(\cdot)$ and $p_{V_{\psi_0}}(\cdot)$ can be expressed as $\ell(\psi_0) - \ell(\psi)$:
\begin{align}
    \textup{KL}(p_{V_{\psi_0}}||p_{V_\psi}) = \int \log p_{V_{\psi_0}}(r) p_{V_{\psi_0}}(r)  \mathrm{d}r - \int \log p_{V_\psi}(r) p_{V_{\psi_0}}(r)  \mathrm{d}r.
\end{align}
The reformulation $\ell(\psi) = \ell(\psi_0) - \textup{KL}(p_{V_{\psi_0}}||p_{V_\psi})$ shows that maximizing the likelihood is equal to minimizing the Kullback-Leibler divergence. \\

\section*{References}
\bibliographystyle{abbrvnat}
\bibliography{references}

\begin{thebibliography}{51}
\providecommand{\natexlab}[1]{#1}
\providecommand{\url}[1]{\texttt{#1}}
\expandafter\ifx\csname urlstyle\endcsname\relax
  \providecommand{\doi}[1]{doi: #1}\else
  \providecommand{\doi}{doi: \begingroup \urlstyle{rm}\Url}\fi

\bibitem[Au and Beck(1999)]{au1999new}
S.-K. Au and J.~L. Beck.
\newblock A new adaptive importance sampling scheme for reliability
  calculations.
\newblock \emph{Structural Safety}, 21\penalty0 (2):\penalty0 135--158, 1999.

\bibitem[Au and Beck(2001)]{au2001estimation}
S.-K. Au and J.~L. Beck.
\newblock Estimation of small failure probabilities in high dimensions by
  subset simulation.
\newblock \emph{Probabilistic Engineering Mechanics}, 16\penalty0 (4):\penalty0
  263--277, 2001.

\bibitem[Au and Wang(2014)]{au2014engineering}
S.-K. Au and Y.~Wang.
\newblock \emph{Engineering Risk Assessment with Subset Simulation}.
\newblock John Wiley \& Sons, 2014.

\bibitem[Beck and Zuev(2015)]{beck2015rare}
J.~L. Beck and K.~M. Zuev.
\newblock Rare event simulation.
\newblock In R.~Ghanem, D.~Higdon, and H.~Owhadi, editors, \emph{Handbook of
  Uncertainty Quantification}, pages 1--26. Springer, Cham, 2015.

\bibitem[Betz et~al.(2014)Betz, Papaioannou, and Straub]{betz2014adaptive}
W.~Betz, I.~Papaioannou, and D.~Straub.
\newblock Adaptive variant of the {BUS} approach to {Bayesian} updating.
\newblock In A.~Cunha, E.~Caetano, P.~Ribeiro, and G.~Müller, editors,
  \emph{Proceedings of the 9th International Conference on Structural Dynamics,
  EURODYN 2014}, pages 3021--3028, 2014.

\bibitem[Bonati et~al.(2019)Bonati, Zhang, and Parrinello]{bonati2019neural}
L.~Bonati, Y.-Y. Zhang, and M.~Parrinello.
\newblock Neural networks-based variationally enhanced sampling.
\newblock \emph{Proceedings of the National Academy of Sciences}, 116\penalty0
  (36):\penalty0 17641--17647, 2019.

\bibitem[Bucklew(2004)]{bucklew2004introduction}
J.~A. Bucklew.
\newblock \emph{Introduction to Rare Event Simulation}, volume~5.
\newblock Springer, New York, 2004.

\bibitem[C{\'e}rou et~al.(2012)C{\'e}rou, Del~Moral, Furon, and
  Guyader]{cerou2012sequential}
F.~C{\'e}rou, P.~Del~Moral, T.~Furon, and A.~Guyader.
\newblock Sequential {Monte Carlo} for rare event estimation.
\newblock \emph{Statistics and Computing}, 22\penalty0 (3):\penalty0 795--808,
  2012.

\bibitem[Chwialkowski et~al.(2016)Chwialkowski, Strathmann, and
  Gretton]{chwialkowski2016kernel}
K.~Chwialkowski, H.~Strathmann, and A.~Gretton.
\newblock A kernel test of goodness of fit.
\newblock In M.~F. Balcan and K.~Q. Weinberger, editors, \emph{International
  Conference on Machine Learning, PMLR}, pages 2606--2615, 2016.

\bibitem[Cotter et~al.(2013)Cotter, Roberts, Stuart, and White]{cotter2013}
S.~L. Cotter, G.~O. Roberts, A.~M. Stuart, and D.~White.
\newblock {MCMC} methods for functions: modifying old algorithms to make them
  faster.
\newblock \emph{Statistical Science}, 28\penalty0 (3):\penalty0 424 -- 446,
  2013.

\bibitem[Goodfellow et~al.(2016)Goodfellow, Bengio, and
  Courville]{goodfellow2016deep}
I.~Goodfellow, Y.~Bengio, and A.~Courville.
\newblock \emph{Deep Learning}.
\newblock MIT press, 2016.

\bibitem[Gorham and Mackey(2015)]{gorham2015measuring}
J.~Gorham and L.~Mackey.
\newblock Measuring sample quality with {Stein}'s method.
\newblock In C.~Cortes, N.~Lawrence, D.~Lee, M.~Sugiyama, and R.~Garnett,
  editors, \emph{Advances in Neural Information Processing Systems 28, NIPS},
  volume~28, 2015.

\bibitem[Gorham and Mackey(2017)]{gorham2017measuring}
J.~Gorham and L.~Mackey.
\newblock Measuring sample quality with kernels.
\newblock In D.~Precup and Y.~W. Teh, editors, \emph{International Conference
  on Machine Learning, PMLR}, pages 1292--1301, 2017.

\bibitem[Hadjidoukas et~al.(2015)Hadjidoukas, Angelikopoulos, Papadimitriou,
  and Koumoutsakos]{hadjidoukas2015pi4u}
P.~E. Hadjidoukas, P.~Angelikopoulos, C.~Papadimitriou, and P.~Koumoutsakos.
\newblock $\pi$4u: A high performance computing framework for {Bayesian}
  uncertainty quantification of complex models.
\newblock \emph{Journal of Computational Physics}, 284:\penalty0 1--21, 2015.

\bibitem[Hasofer and Lind(1974)]{hasofer1974exact}
A.~M. Hasofer and N.~C. Lind.
\newblock Exact and invariant second-moment code format.
\newblock \emph{Journal of the Engineering Mechanics Division}, 100\penalty0
  (1):\penalty0 111--121, 1974.

\bibitem[Hastings(1970)]{hastings1970monte}
W.~K. Hastings.
\newblock {{Monte Carlo sampling methods using Markov chains and their
  applications}}.
\newblock \emph{Biometrika}, 57\penalty0 (1):\penalty0 97--109, 04 1970.

\bibitem[Hohenbichler and Rackwitz(1988)]{hohenbichler1988improvement}
M.~Hohenbichler and R.~Rackwitz.
\newblock Improvement of second-order reliability estimates by importance
  sampling.
\newblock \emph{Journal of Engineering Mechanics}, 114\penalty0 (12):\penalty0
  2195--2199, 1988.

\bibitem[Invernizzi et~al.(2017)Invernizzi, Valsson, and
  Parrinello]{invernizzi2017coarse}
M.~Invernizzi, O.~Valsson, and M.~Parrinello.
\newblock Coarse graining from variationally enhanced sampling applied to the
  {Ginzburg}--{Landau} model.
\newblock \emph{Proceedings of the National Academy of Sciences}, 114\penalty0
  (13):\penalty0 3370--3374, 2017.

\bibitem[Jensen et~al.(2013)Jensen, Vergara, Papadimitriou, and
  Millas]{jensen2013use}
H.~Jensen, C.~Vergara, C.~Papadimitriou, and E.~Millas.
\newblock The use of updated robust reliability measures in stochastic
  dynamical systems.
\newblock \emph{Computer Methods in Applied Mechanics and Engineering},
  267:\penalty0 293--317, 2013.

\bibitem[Juneja and Shahabuddin(2006)]{juneja2006rare}
S.~Juneja and P.~Shahabuddin.
\newblock Rare-event simulation techniques: an introduction and recent
  advances.
\newblock \emph{Handbooks in Operations Research and Management Science},
  13:\penalty0 291--350, 2006.

\bibitem[Kahn and Marshall(1953)]{kahn1953methods}
H.~Kahn and A.~W. Marshall.
\newblock Methods of reducing sample size in {Monte Carlo} computations.
\newblock \emph{Journal of the Operations Research Society of America},
  1\penalty0 (5):\penalty0 263--278, 1953.

\bibitem[Kanjilal et~al.(2023)Kanjilal, Papaioannou, and
  Straub]{kanjilal2023bayesian}
O.~Kanjilal, I.~Papaioannou, and D.~Straub.
\newblock {Bayesian} updating of reliability by cross entropy-based importance
  sampling.
\newblock \emph{Structural Safety}, 102:\penalty0 102325, 2023.

\bibitem[K{\"a}stner and Thiel(2005)]{kastner2005bridging}
J.~K{\"a}stner and W.~Thiel.
\newblock Bridging the gap between thermodynamic integration and umbrella
  sampling provides a novel analysis method:“{Umbrella} integration”.
\newblock \emph{The Journal of Chemical Physics}, 123\penalty0 (14):\penalty0
  144104, 2005.

\bibitem[Kingma and Ba(2015)]{kingma2014adam}
D.~P. Kingma and J.~Ba.
\newblock Adam: A method for stochastic optimization.
\newblock In Y.~Bengio and Y.~LeCun, editors, \emph{3rd International
  Conference for Learning Representations, ICLR, Conference Track Proceedings},
  2015.

\bibitem[Koutsourelakis et~al.(2004)Koutsourelakis, Pradlwarter, and
  Schueller]{koutsourelakis2004reliability}
P.-S. Koutsourelakis, H.~J. Pradlwarter, and G.~I. Schueller.
\newblock Reliability of structures in high dimensions, part i: algorithms and
  applications.
\newblock \emph{Probabilistic Engineering Mechanics}, 19\penalty0 (4):\penalty0
  409--417, 2004.

\bibitem[Lelièvre et~al.(2010)Lelièvre, Stoltz, and Rousset]{stoltz2010free}
T.~Lelièvre, G.~Stoltz, and M.~Rousset.
\newblock \emph{Free Energy Computations: A Mathematical Perspective}.
\newblock World Scientific, 2010.

\bibitem[Liu et~al.(2016)Liu, Lee, and Jordan]{liu2016kernelized}
Q.~Liu, J.~Lee, and M.~Jordan.
\newblock A kernelized {Stein} discrepancy for goodness-of-fit tests.
\newblock In M.~F. Balcan and K.~Q. Weinberger, editors, \emph{International
  Conference on Machine Learning, PMLR}, pages 276--284, 2016.

\bibitem[Liu et~al.(2020)Liu, Gao, and Yin]{liu2020improved}
Y.~Liu, Y.~Gao, and W.~Yin.
\newblock An improved analysis of stochastic gradient descent with momentum.
\newblock \emph{Advances in Neural Information Processing Systems},
  33:\penalty0 18261--18271, 2020.

\bibitem[McMahon(2018)]{mcmahon2018linear}
S.~J. McMahon.
\newblock The linear quadratic model: usage, interpretation and challenges.
\newblock \emph{Physics in Medicine \& Biology}, 64\penalty0 (1):\penalty0
  01TR01, 2018.

\bibitem[Metropolis et~al.(1953)Metropolis, Rosenbluth, Rosenbluth, Teller, and
  Teller]{metropolis1953equation}
N.~Metropolis, A.~W. Rosenbluth, M.~N. Rosenbluth, A.~H. Teller, and E.~Teller.
\newblock Equation of state calculations by fast computing machines.
\newblock \emph{The Journal of Chemical Physics}, 21\penalty0 (6):\penalty0
  1087--1092, 1953.

\bibitem[Owen and Zhou(2000)]{owen2000safe}
A.~Owen and Y.~Zhou.
\newblock Safe and effective importance sampling.
\newblock \emph{Journal of the American Statistical Association}, 95\penalty0
  (449):\penalty0 135--143, 2000.

\bibitem[Papadimitriou et~al.(2001)Papadimitriou, Beck, and
  Katafygiotis]{papadimitriou2001updating}
C.~Papadimitriou, J.~L. Beck, and L.~S. Katafygiotis.
\newblock Updating robust reliability using structural test data.
\newblock \emph{Probabilistic Engineering Mechanics}, 16\penalty0 (2):\penalty0
  103--113, 2001.

\bibitem[{R Core Team}(2021)]{Rstats}
{R Core Team}.
\newblock \emph{R: A Language and Environment for Statistical Computing}.
\newblock R Foundation for Statistical Computing, Vienna, Austria, 2021.
\newblock URL \url{https://www.R-project.org/}.

\bibitem[Robert et~al.(2018)Robert, Elvira, Tawn, and
  Wu]{robert2018accelerating}
C.~P. Robert, V.~Elvira, N.~Tawn, and C.~Wu.
\newblock Accelerating {MCMC} algorithms.
\newblock \emph{Wiley Interdisciplinary Reviews: Computational Statistics},
  10\penalty0 (5):\penalty0 e1435, 2018.

\bibitem[Rubino and Tuffin(2009)]{rubino2009rare}
G.~Rubino and B.~Tuffin.
\newblock \emph{Rare Event Simulation Using {Monte Carlo} Methods}.
\newblock John Wiley \& Sons, 2009.

\bibitem[Sch{\"o}bi et~al.(2017)Sch{\"o}bi, Sudret, and
  Marelli]{schobi2017rare}
R.~Sch{\"o}bi, B.~Sudret, and S.~Marelli.
\newblock Rare event estimation using polynomial-chaos kriging.
\newblock \emph{ASCE-ASME Journal of Risk and Uncertainty in Engineering
  Systems, Part A: Civil Engineering}, 3\penalty0 (2):\penalty0 D4016002, 2017.

\bibitem[Scott(2015)]{scott2015multivariate}
D.~W. Scott.
\newblock \emph{Multivariate Density Estimation: Theory, Practice, and
  Visualization}.
\newblock John Wiley \& Sons, 2015.

\bibitem[Shao(2010)]{shao2010dependent}
X.~Shao.
\newblock The dependent wild bootstrap.
\newblock \emph{Journal of the American Statistical Association}, 105\penalty0
  (489):\penalty0 218--235, 2010.

\bibitem[Shirts and Ferguson(2020)]{shirts2020statistically}
M.~R. Shirts and A.~L. Ferguson.
\newblock Statistically optimal continuous free energy surfaces from biased
  simulations and multistate reweighting.
\newblock \emph{Journal of Chemical Theory and Computation}, 16\penalty0
  (7):\penalty0 4107--4125, 2020.

\bibitem[Song and Kingma(2021)]{song2021train}
Y.~Song and D.~P. Kingma.
\newblock How to train your energy-based models.
\newblock \emph{arXiv preprint arXiv:2101.03288}, 2021.

\bibitem[Stecher et~al.(2014)Stecher, Bernstein, and
  Cs{\'a}nyi]{stecher2014free}
T.~Stecher, N.~Bernstein, and G.~Cs{\'a}nyi.
\newblock Free energy surface reconstruction from umbrella samples using
  {Gaussian} process regression.
\newblock \emph{Journal of Chemical Theory and Computation}, 10\penalty0
  (9):\penalty0 4079--4097, 2014.

\bibitem[Stein(1972)]{stein1972bound}
C.~Stein.
\newblock A bound for the error in the normal approximation to the distribution
  of a sum of dependent random variables.
\newblock In \emph{Proceedings of the Sixth Berkeley Symposium on Mathematical
  Statistics and Probability, Volume 2: Probability Theory}, pages 583--602,
  1972.

\bibitem[Straub(2011)]{straub2011reliability}
D.~Straub.
\newblock Reliability updating with equality information.
\newblock \emph{Probabilistic Engineering Mechanics}, 26\penalty0 (2):\penalty0
  254--258, 2011.

\bibitem[Straub and Papaioannou(2015)]{straub2015bayesian}
D.~Straub and I.~Papaioannou.
\newblock {Bayesian} updating with structural reliability methods.
\newblock \emph{Journal of Engineering Mechanics}, 141\penalty0 (3):\penalty0
  04014134, 2015.

\bibitem[Straub et~al.(2016)Straub, Papaioannou, and Betz]{straub2016bayesian}
D.~Straub, I.~Papaioannou, and W.~Betz.
\newblock {Bayesian} analysis of rare events.
\newblock \emph{Journal of Computational Physics}, 314:\penalty0 538--556,
  2016.

\bibitem[Sundar and Manohar(2013)]{sundar2013updating}
V.~Sundar and C.~Manohar.
\newblock Updating reliability models of statically loaded instrumented
  structures.
\newblock \emph{Structural Safety}, 40:\penalty0 21--30, 2013.

\bibitem[Tokdar and Kass(2010)]{tokdar2010importance}
S.~T. Tokdar and R.~E. Kass.
\newblock Importance sampling: a review.
\newblock \emph{Wiley Interdisciplinary Reviews: Computational Statistics},
  2\penalty0 (1):\penalty0 54--60, 2010.

\bibitem[Torrie and Valleau(1977)]{torrie1977nonphysical}
G.~M. Torrie and J.~P. Valleau.
\newblock Nonphysical sampling distributions in {Monte Carlo} free-energy
  estimation: Umbrella sampling.
\newblock \emph{Journal of Computational Physics}, 23\penalty0 (2):\penalty0
  187--199, 1977.

\bibitem[Valsson and Parrinello(2014)]{valsson2014variational}
O.~Valsson and M.~Parrinello.
\newblock Variational approach to enhanced sampling and free energy
  calculations.
\newblock \emph{Physical Review Letters}, 113\penalty0 (9):\penalty0 090601,
  2014.

\bibitem[Van~der Vaart(2000)]{van2000asymptotic}
A.~W. Van~der Vaart.
\newblock \emph{Asymptotic Statistics}, volume~3.
\newblock Cambridge University Press, 2000.

\bibitem[Wang and Landau(2001)]{wang2001efficient}
F.~Wang and D.~P. Landau.
\newblock Efficient, multiple-range random walk algorithm to calculate the
  density of states.
\newblock \emph{Physical Review Letters}, 86\penalty0 (10):\penalty0 2050,
  2001.

\end{thebibliography}

\end{document}